\documentclass[9pt,  twoside]{pnas-new}

\usepackage{amsfonts}
\usepackage{amsmath}
\usepackage{enumerate}
\usepackage{bbm,color}

\newcommand{\rmd}{\mathrm{d}}

\newcommand{\ab}{\ensuremath{\mathbf{b}}}
\newcommand{\rv}{\ensuremath{\mathbf{r}}}
\newcommand{\xv}{\ensuremath{\mathbf{x}}}
\newcommand{\yv}{\ensuremath{\mathbf{y}}}

\newcommand{\nullv}{\ensuremath{\mathbf{0}}}
\newcommand{\I}{\ensuremath{\mathrm{I}}}
\newcommand{\Y}{\ensuremath{\mathrm{Y}}}
\newcommand{\Ph}{\ensuremath{\mathrm{P}}}

\newcommand{\lambdav}{   \ensuremath{\boldsymbol{\lambda}}         }
\newcommand{\1}{\ensuremath{\mathbf{1}}}
\newcommand{\B}{\ensuremath{\mathbf{B}}}
\newcommand{\Ev}{\ensuremath{   \boldsymbol{\mathcal{E}}}}
\newcommand{\E}{\ensuremath{  \mathcal{E}           }}
\newcommand{\Lagr}{\ensuremath{  \mathcal{L}           }}
\newcommand{\SI} {\emph{SI}} 
\newcommand{\MM} {\emph{Materials and Methods}} 
\usepackage{color}

\newcommand{\Rmatrix}{\ensuremath{\mathbf{R}}}

\templatetype{pnasresearcharticle} 

\title{Roles of repertoire diversity in robustness of humoral immune response}

\author[1,\dag]{Alexander Mozeika}
\author[2]{Franca Fraternali} 
\author[3]{Deborah Dunn-Walters}
\author[1,2,4]{Anthony C. C. Coolen}

\affil[1]{Institute for Mathematical and Molecular Biomedicine, King's College London, Hodgkin Building, London, UK}
\affil[2]{Randall Division of Cell and Molecular Biophysics, School of Basic and Medical Biosciences, King's College London,
London, UK}
\affil[3]{Faculty of Health and Medical Sciences, School of Biosciences and Medicine, University of Surrey, Guildford, Surrey, UK}
 \affil[4]{Department of Mathematics, King's College London, The Strand, London, UK}

\leadauthor{Alexander Mozeika} 

\significancestatement{The older immune system is less able to protect us from infection and more likely to malfunction, and inappropriate inflammation is involved in the aetiology of many diseases of old age.  Since the world population is growing older, immune senescence is a significant health risk.  Previous studies, by us and others, show that the human antibody repertoire is less diverse and there are more antibodies that recognise self-antigens in older people.  We posed the scenario that an antibody can bind multiple different targets, both self and non-self, but with varying affinity, and asked how efficacy of the immune system might be affected by this balance and by the loss of diversity of antibodies at a population level. Our theoretical framework was developed from first principles.  It predicts that a reduced diversity and increased self-reactivity in the antibody pool will slow down immune responses to exogenous targets, thus providing an explanation for the reduced immune response to vaccines and infections in older people.}

\authorcontributions{A.M., F.F., D.D.-W. and  A.C.C.C. designed research, performed research  and wrote the paper.}
\authordeclaration{The authors declare no conflict of interest.}
\correspondingauthor{\textsuperscript{\dag}To whom correspondence should be addressed. E-mail: alexander.mozeika@kcl.ac.uk}

\keywords{adaptive immune response $|$ immune repertoire $|$  repertoire diversity  $|$   repertoire self-reactivity} 

\begin{abstract}
The adaptive immune system relies on diversity  of its repertoire of receptors to protect the organism from a great variety of pathogens.  Since the  initial repertoire is the  result of random gene rearrangement, binding of receptors is not limited to pathogen-associated antigens but also includes self antigens. There is a fine balance between having a diverse repertoire, protecting  from  many different pathogens,  and yet reducing its self-reactivity as far as possible to avoid damage  to self. In the ageing immune system this  balance  is altered, manifesting in reduced specificity of response to pathogens or  vaccination on a background of higher self-reactivity. To answer the question whether age-related changes of repertoire in the diversity and self/non-self affinity balance of antibodies could explain the reduced efficacy of the humoral response in older people,  we construct  a minimal  mathematical model  of the humoral immune response. The principle of least damage  allows us, for a given repertoire of antibodies, to resolve a   tension between the necessity to neutralise target antigens as quickly as possible and the requirement to limit the damage to self antigens  leading to an optimal  dynamics of immune response.  The model  predicts slowing down of  immune response  for repertoires  with  reduced diversity  and increased self-reactivity. 
\end{abstract}

\dates{This manuscript was compiled on \today}
\doi{\url{www.pnas.org/cgi/doi/10.1073/pnas.XXXXXXXXXX}}

\begin{document}

\maketitle
\thispagestyle{firststyle}
\ifthenelse{\boolean{shortarticle}}{\ifthenelse{\boolean{singlecolumn}}{\abscontentformatted}{\abscontent}}{}


\dropcap{T}he adaptive immune system relies on an extremely diverse repertoire of receptors that can recognise target molecules to protect us from pathogens.  Each cell has a unique specificity, encoded by the T cell receptor on T cells, or the B cell receptor on B cells.  In the case of B cells, the B cell receptor is also known as surface immunoglobulin, and this immunoglobulin (Ig) can be secreted as antibody once the cell has developed into a plasma cell. Antibodies (Ab) are an important first line of defence, they can block the action of harmful target molecules and help to recruit additional elements of the immune system by acting as bridges between target molecules and effector cells. The targets of Ab are known as antigens (Ag).

B cells are formed in the bone marrow, where they acquire a unique Ig via gene rearrangement, a process that can produce over $10^8$ different genes by reassortment of less than 200 germline gene segments~\cite{Dunn-Walters2018,Dunn-Walters2016}. The highest diversity is seen in the areas of the Ig gene where different gene segments are joined together, and these areas of the gene encode the parts of the Ab that bind to Ag, thus ensuring a large  diversity in the Abs structural forms of possible binding interactions~\cite{Dondelinger2018}.  Since gene rearrangement is essentially random, the potential binding interactions of the initial repertoire are not limited to pathogen-associated target Ag, they can include self-Ag also.  Immunological tolerance is a negative selection process whereby B cells having Ig with strong binding to self are deleted from the repertoire so that they cannot develop into plasma cells secreting self-reactive Abs~\cite{Martin2016}.  There is a trade-off between having a large enough shape space to be prepared for many different pathogen-associated Ags and yet reducing self-reactivity as far as possible to avoid self-damage~\cite{Childs2015}. During activation of B cells in an immune response, the B cells with specificity for target Ag are expanded~\cite{Wu2012}.   With the advent of high throughput sequencing methods, we can see that there are a broad range of antibodies that respond, even for simple antigens such as tetanus toxin~\cite{Poulsen2007}.  The affinity for target Ag can be increased in germinal centres of secondary lymphoid tissue where B cells undergo cycles of somatic hypermutation of their Ig genes, followed by competitive selection for the best target Ag-binders~\cite{Bannard2017, Ademokun2011}.  Thus, the initial repertoire is altered by both positive and negative selection events, depending on binding to target and self Ags.

Older people are more susceptible to infection, in particular to bacterial infections such as pneumonia or urinary tract infections~\cite{Dunn-Walters2016}.  In the ageing immune system, the balance of the immune system is altered, manifested in a reduced specific target Ab response to infection or vaccination on a background of a higher number of Abs showing evidence of self-reactivity~\cite{Bannard2017}.  In this instance, the presence of self-reactive Abs does not usually indicate autoimmune disease pathology, rather we believe it may reflect an increased presence of `polyspecific' or `promiscuous'  antibodies which have binding affinities that are measurable for several different targets.  Since we know that T cell availability and function is also compromised with age~\cite{Goronzy2019}, it is possible that the B cell repertoire is not receiving as much help to produce affinity-matured specific antibodies that can dominate the immune response, relying instead on more T-independent responses.  Increased use of IgG2 over IgG1  detected in the samples of older patients supports this hypothesis~\cite{Martin2015}.  Analyses of older Ig gene repertoires indicate that selection events at different stages of B cell development, both positive and negative, are less effective in the older immune system~\cite{Dunn-Walters2016}.  Some Ig gene characteristics that have been associated with polyspecificity are seen to be increased in the  na\"ive B cell population of older people~\cite{Laffy2017}.  In addition, a reduction in the diversity of the B cell repertoire overall has also been seen in older people~\cite{Gibson2009}.  

Our question is whether age-related repertoire changes in diversity and target/self-Ag affinity balance could explain the reduced efficacy of the humoral response in older people.  To this end we construct  a minimal  mathematical model  of the humoral immune response.  The  ingredients of this model are Abs, target Ag and  self-Ag.   Abs are binding the  target Ag and thus reduce  the  amount of  free  target  Ag, i.e. Ag not bound by Abs.  The amount of free target  Ag plays a  role of an `energy'  in our  construction, and we assume that the immune system tries to minimise  this energy.  We note that various energy functions have been used in immune system modelling in the past, such as  the `total affinity'  in somatic hypermutation of B cells~\cite{Kepler1993},  or the `disagreement'  between the B and T cell  signalling in lymphocyte `networks'  in more recent studies~\cite{Agliari2011, Bartolucci2015, Mozeika2016}.

Furthermore, we assume that we have many types of Abs, each  specified by its affinity to the  targets and to self Ag~\cite{Theofilopoulos2017}, which  constitute the immune \emph{repertoire} in our model.  Immune repertoires were studied theoretically in e.g.~\cite{Perelson1979, DeBoer1993}, and more recently in~\cite{Mayer2015}.  The role of self-Ags in shaping the  \emph{diversity} of repertoires, important for  reliable self/non-self discrimination~\cite{Perelson1979},  was emphasised in~\cite{DeBoer1993}.  We assume that both  the binding of Abs  to self-Ag and the presence of free target Ag  incurs  \emph{damage}, hence the \emph{unconstrained} use of Abs is not possible and the amount of free target Ag has to be reduced.  To resolve these two conflicting requirements we develop the \emph{principle of least damage} which allows us to derive an \emph{optimal}  dynamics of the immune response. While the resulting theoretical framework is very general,  even  its simplest analytically solvable version   predicts the  `slowing down' of  the immune response  for repertoires  with  reduced diversity and increased self-reactivity.

\section*{Mechanics of Immune Response\label{section:mechanics}}

\subsection*{A simple thought experiment\label{ssection:experiment}}
\begin{figure}
\centering
\includegraphics[width=.65\linewidth]{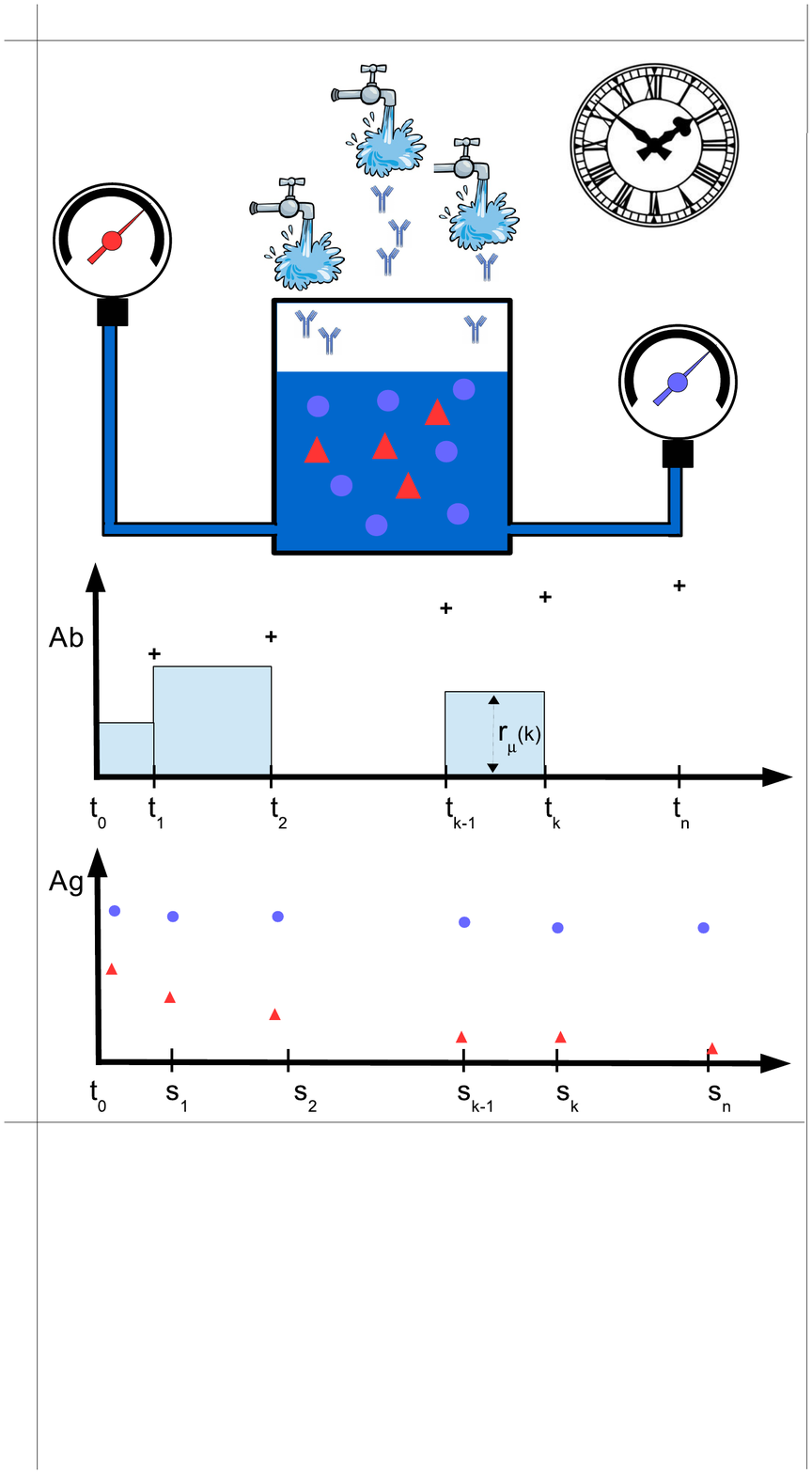}
\caption{Immune Response: the Thought Experiment. \emph{Top drawing}:  antibodies (Abs) are released into a reservoir  which contains a mixture  of  target antigen, Ag (red triangles) and self-antigen, self-Ag (blue circles). They can form Ab-Ag complexes and thereby reduce  the amount of \emph{free} (i.e. unbound) Abs, target Ag and self-Ag. The latter two amounts are measured, respectively,  by the left  and right  `gauges'.  The experiment is performed under  \emph{constraints},  such as finite duration and  finite reservoir volume.  \emph{Middle drawing:} the release  of antibodies is controlled by the  flow rate (vertical axis) at any given  time (horizontal axis). The total amount of Ab released up to time $t_k$ (crosses) is increasing with time.  \emph{Bottom drawing:} the amount of free target  Ag (self-Ag) is decreasing  with time.  Each measurement is  taken  at the time-point  $s_k$ with   $s_k \gg t_k$,   to ensure that the mixture in  the reservoir is always in equilibrium.
}
 \label{figure:experiment} 
\end{figure}

To investigate the trade-off between antibody binding to a desired target, such as pathogen, versus a self-damaging target, we consider the case where there are many antibodies responding to a challenge, in the absence of a single dominating high-affinity antibody.  Our thought experiment assumes that we have a finite volume reservoir containing a finite amount of target antigen (Ag)  and self-antigen (self-Ag)  in  some medium (see Figure \ref{figure:experiment}).  We also assume that we are {given}   $M$ different types  of antibodies (Abs), labelled by the integers  $1$ to $M$,  which can be released  into the reservoir. The release of each Ab is controlled   by a  valve.   We assume that  the reservoir contents are  well mixed.  Abs  released into the reservoir  react  with both  types of Ag, resulting in the formation of  Ag-Ab complexes; thus  the amount of `free' (i.e. unbound)  Ag is reduced.  The properties of Abs,  such as how strongly they react with each Ag, etc.,    are assumed to be initially  {unknown}. Two gauges attached to the reservoir  measure the amounts of  free  target Ag and  of self-Ag.  The opening and closing of valves, and performing various measurements (such as of the amount of Abs delivered   into the reservoir,  the amount of  free target Ag and  self-Ag in the reservoir) constitutes an `experiment'.  
\subsection*{Measurement protocol}
 The experimental measurement is defined by a set of time points $t_0,\ldots, t_{k-1}, t_k, \ldots, t_n$ together with the flow rates   $r_\mu(t_1), \ldots, r_\mu(t_{k-1}), r_\mu(t_{k}), \dots, r_\mu(t_n)$ recorded at these times, for each Ab $\mu$   (see  Figure \ref{figure:experiment}). We label antibody types by Greek indices. The total amount of Ab $\mu$ released into the reservoir up to the time $t_k$ is given by  the sum  $b_\mu(t_k)=\sum_{\ell=1}^k r_\mu(t_{\ell})(t_\ell- t_{\ell-1})$. If the flow rates $r_\mu(t)$ are smooth functions of time, each amount approaches an integral $b_\mu(t_k)=\int_{t_0}^{t_k} r_\mu(t)\rmd t$ in the limit where the measurement times become arbitrarily close,   $t_\ell- t_{\ell-1}\rightarrow0$.  The system in  Figure \ref{figure:experiment} is then fully described by the amounts of Abs $\ab(t)=(b_1(t), \dots,b_M(t))$,  delivered into the reservoir up to time $t$, and the rates  $\frac{\rmd}{\rmd t}\ab(t)=\left(\frac{\rmd}{\rmd t}b_1(t), \dots,\frac{\rmd}{\rmd t} b_M(t)\right)$ of delivery of Abs.   The amount of free target Ag, measured by the left gauge in Figure \ref{figure:experiment},  is a function $A_T(\ab(t))$  of the Abs  $\ab(t)$. The same is true for  $A_S(\ab)$, the amount of free self-Ag, measured by the right gauge in  the Figure \ref{figure:experiment}. By construction, the total amount of free Ag in the experiment is a non-increasing function of time, i.e. $\frac{\rmd}{\rmd t} A_T\leq0$ and $\frac{\rmd}{\rmd t} A_S\leq0$. 

\subsection*{Measurement of antibody affinity\label{ssection:single-Ab-experiment}}
Let the amount of free target Ag at time $t_{0}$ be $A_T(\ab(t_{0}))$, and assume that at the next time-point $t_{1}$ we release into the reservoir a small amount $\Delta b_\mu$ of Ab $\mu$, i.e. $b_\mu(t_{1})=b_\mu(t_{0})+\Delta b_\mu$ and $b_\nu(t_{1})=b_\nu(t_{0})$ for all $\nu\neq\mu$.  The resulting  change in the amount of free target Ag is given  by  $\Delta A_T^{\mu}=A_T(\ab(t_{1}))-A_T(\ab(t_{0}))\leq0$ and for $\Delta b_\mu\rightarrow0$ we have  $(\partial A_T/\partial b_\mu)(\rmd b_\mu/\rmd t)\leq0$.  The same holds for   the free self-Ag $A_S(\ab)$.  
Upon releasing  a single Ab into the reservoir we will generally observe different behaviours of the gauges,  which can be used to classify this Ab.  Ab  $\mu$ is more `reactive' than  Ab  $\nu$ if $\Delta A_T^{\mu}\leq\Delta A_T^{\nu}$, for $\Delta b_\mu=\Delta b_\nu$, i.e. if the same amount of Ab reduces more  Ag upon releasing type $\mu$ insterad of $\nu$ .  Similarly, Ab $\mu$ is  more {self-reactive} than  Ab $\nu$ when  $\Delta A_S^{\mu}\leq\Delta A_S^{\nu}$, and Ab $\mu$ is more reactive than self-reactive when $\Delta A_T^{\mu}\leq\Delta A_S^{\mu}$ (and vice versa).  For  $\Delta b_\mu\rightarrow0$ all of the above definitions can implemented with partial derivatives, so Ab $\mu$ is more reactive than self-reactive when  $(\partial A_T /\partial b_\mu) \leq (\partial A_S /\partial b_\mu )$, etc. 

The difference $\Delta A_T^{\mu}$ is related to the \emph{affinity} of Ab $\mu$~\cite{Janeway2012}, which is usually defined as the ratio  $r_\mu=K_{\mu}^{+}/K_{\mu}^{-}$ of forward/backward  rates of the chemical reaction  $Ag + Ab  \rightleftharpoons AgAb$. In chemical equilibrium  the latter can be computed  experimentally, via the relation $r_\mu=[AgAb]/[Ag] [Ab]$,  upon measuring the amount $[Ag]$ of free target Ag, the amount $[Ab]$ of free Ab, and the amount $[AgAb]$ of Ag-Ab complexes, in the absence of other antibodies or antigens.
In our notation, the affinity can be written as 
\begin{eqnarray}
r_\mu= -\frac{\left([Ag]\!-\![AgAb]\right)\! -\! [Ag] }{ [Ab]\!-\!0}\frac{1}{[Ag]}=-\frac{\Delta A_T^{\mu}}{\Delta b_\mu}\frac{1}{A_T(\nullv)},
\end{eqnarray} 
evaluated at $\ab=\nullv$.
Thus for $\Delta b_\mu\rightarrow0$ it becomes the  derivative 
\begin{eqnarray}
r_\mu(\ab)&=&-\Big(\frac{\partial   }{\partial b_\mu}\log A_T (\ab)\Big)_{\ab=\nullv}
\label{def:affinity}. 
\end{eqnarray}
   For $\ab\neq\nullv$, expression [\ref{def:affinity}] can be seen as a \emph{generalised   affinity},  measured  by adding a small amount of Ab $\mu$ in to the  \emph{mixture} of  Ags and  Abs.  The affinity to self-Ag $r^S_\mu(\ab)$ uses the same definition as [\ref{def:affinity}],  but with  $A_S (\ab)$ instead of $A_T (\ab)$.

In immunology one commonly thinks in terms of a repertoire of  different antibodies, each reacting to target-Ag or to self-Ag, and of changing repertoires representing expansions of target-Ag antibodies in immune activation and deletion of self-Ag antibodies in immune tolerance.  However, single antibodies can bind to multiple different antigens, with varying affinity, and these antigens could be either target-Ag or self-Ag. What we may have empirically determined to be a specific target-Ag binding antibody may in fact be a polyspecific antibody where the binding to self-Ag is so small as to be unnoticed. So we need to consider polyspecific antibodies,  with variable affinities for binding to multiple Ag.

\subsection*{Using  multiple antibody types to reduce free antigen\label{ssection:multiple-Ab-experiment}}
We assume here for simplicity that we have one type of target Ag, which we seek to reduce using a repertoire of antibodies.  The Ag has $N_A$ distinct regions which can be `recognised' by Abs,  the \emph{epitopes}. The Abs,  represented by the amounts  $\ab=(b_1, \ldots, b_M)$, are assumed to interact with free  epitopes, i.e. those not bound by Abs. The amounts of the free epitopes are written as $\Ev=(\E_1,\ldots,\E_{N_A})$.  
Each $\E_i\equiv\E_i(\ab)$ must be a non-decreasing function of the amount of Abs, such that  $0\leq\E_i(\ab)\leq\E_i(\nullv)$.  Furthermore, the `amount' of free target antigen $A_T(\ab)\equiv A_T\left(\Ev\left(\ab\right)\right)\geq0$ will similarly be  a non-decreasing function  of the amount of free epitopes. 

We assume that the protocol used to reduce the amount of Ag takes  the form of differential equations for the rates of antibody delivery, given the  amounts $\ab\equiv \ab (t)$  of Abs in the reservoir (as in biological processes), i.e. that
 \begin{eqnarray}
\frac{\rmd}{\rmd t}{b}_\mu&=& f_\mu\left( \ab  \right) \label{eq:dbdt-1} 
\end{eqnarray}
For the dynamics [\ref{eq:dbdt-1}] to reduce target Ag, it is sufficient 
that  the rate functions $f_\mu\!\left( \ab  \right)$ are positive,
 \begin{eqnarray}
~~\frac{\rmd  }{\rmd t}A_T=  \sum_{\mu=1}^M    \frac{\partial A_T}{\partial   b_\mu}\frac{\rmd}{\rmd t}{b}_\mu = -A_T(\ab)\!\!\sum_{\mu=1}^M r_\mu(\ab)  f_\mu\left( \ab \right)\leq0. 
\label{eq:dAdt}
\end{eqnarray}
Clearly, since $A_T(\ab)\geq0$, the $A_T(\ab)$  is a Lyapunov function of [\ref{eq:dbdt-1}].  The possible choices for  the Ab delivery rate  functions $f_\mu\!\left( \ab  \right)$  are further restricted  by physical constraints in the experiment, such as finite  time,  finite volume,   finite amount of available Abs,  etc.   
Further complications occur if,   in addition to target Ag, the reservoir also contains self Ag and,  when we try to reduce free target Ag,  only a finite amount of reduced self Ag (off-target damage) can be tolerated. It is natural to assume that  the amount of free self Ag must depend in a similar way on the amount of free  epitopes $\Ev^S(\ab)= \left(\E_1^S(\ab),\ldots, \E_{N_S}^S(\ab)\right)$ as the target antigen, so $A_S(\ab)=A_S\left(\Ev^S(\ab)\right)$. Furthermore, one would expect that the Ab dynamics [\ref{eq:dbdt-1}] is also a function of self-epitopes, i.e.
 \begin{eqnarray}
\frac{\rmd}{\rmd t}{b}_\mu&=& f_\mu\left( \Ev(\ab),  \Ev^S(\ab) \right)\!,  \label{eq:dbdt-2}
\end{eqnarray}
and that any biologically sensible choice $f_\mu(\ldots)$ must be an {increasing} function of 
 $\Ev(\ab)$ and a {decreasing} function of $\Ev^S(\ab)$.  

\section*{Antibody Dynamics}

\subsection*{Principle of least damage\label{section:damage}}%

Instead of guessing  an equation for the Ab delivery rates $f_\mu(\ldots)$, we take a Darwinian approach and assume that an optimized mechanism will have evolved that reduces the target Ag \emph{as quickly as possible},  to minimise the `damage' done,  while minimising the harmful binding to self Ag in the process. The optimization problem can be solved using mathematical tools from physics. To this end we  consider all possible  \emph{paths}  $\ab(t)$, allowed by the setup   in Figure \ref{figure:experiment}.  Any such path will obey $\rmd b_\mu/\rmd t\geq0$ and ${\rmd} A_T/\rmd t\leq 0$, i.e. each will minimize  $A_T(\ab)$ (which we will call the `potential energy'). The latter is  a property  of  the reservoir. We assume  that the antibody delivery mechanism in Figure \ref{figure:experiment} has associated with it  a `kinetic  energy'  $\mathcal{T}(\rmd{\ab}/\rmd t)$, which reflects the likely involvement of further variables governed by first order differential equations (equivalently, that the equations for $b_\mu$, if autonomous, will be at least second order).   The path which begins at $\ab(t_0)$ at time $t_0$ and ends in  $\ab(t_1)$  at time $t_1>t_0$, with $A_{T}(\ab(t_0))\geq A_T(\ab(t_1))$,  can then be obtained~\cite{Arnold1989} by minimising the \emph{action}
\begin{eqnarray}
\mathcal{S}\left(\ab, \frac{\rmd}{\rmd t}{\ab}\right)=\int_{t_0}^{t_1}\!\rmd t~ \Lagr\left(\ab(t), \frac{\rmd}{\rmd t}{\ab}(t)  \right) \label{def:S},
\end{eqnarray}
where  $\Lagr\left(\ab,\frac{\rmd}{\rmd t}{\ab}  \right)=  A_T(\ab)-\mathcal{T}(\frac{\rmd}{\rmd t}{\ab})$ is  the Lagrangian (see \emph{Materials and Methods}). 

\subsection*{Interpretation of the action}

The area under the curve of $A_T(\ab(t))$ on any path $\ab(t)$, given by the integral 
\begin{eqnarray}
\mathcal{D}_A\left(t_1-t_0\right)&=&\int_{t_0}^{t_1}A_T(\ab(t))\,\rmd t\label{def:damage},
\end{eqnarray}
can be seen as a damage inflicted upon the organism during the time interval $[t_0, t_1]$ by the presence of free target Ag.  The intuition is that during any  small  time interval  the damage inflicted by Ag is equal to   the amount of free  Ag  times the  time it spends in the organism.  Definition  [\ref{def:damage}] assumes moreover  that this damage is \emph{cumulative}, i.e. exposure  to a  large amount of Ag for a short time or a to a small  amount of Ag  for  a longe time are equivalent.  We observe that  $0\leq\mathcal{D}_A\leq A_T(\ab(t_0))\left(t_1-t_0\right)$, which follows from the properties $A_T(\ab(t))\geq 0$ and $A_T(\ab(t_0))\geq A_T(\ab(t_1))$.  So the path  minimising the action [\ref{def:S}] is the path which minimises the damage $\mathcal{D}_A\left(t_1-t_0\right)$, but subject to the constraint on $\rmd{\ab}/\rmd t$  enforced by the term  $\int_{t_0}^{t_1}\!\rmd t~ \mathcal{T}\left(\frac{\rmd}{\rmd t } \ab(t) \right)$ in the action \cite{Gelfand2000}.  

Similar to  [\ref{def:damage}], we can consider  the integral 
\begin{eqnarray}
\mathcal{D}_S\left(t_1-t_0\right)=\int_{t_0}^{t_1}\!\rmd t~A_S(\ab(t))\,\label{def:self-damage},
\end{eqnarray}
where $0\leq\mathcal{D}_S\leq A_S(\ab(t_0))\left(t_1-t_0\right)$. From this integral follows the  `damage to self', defined for each small time interval as the amount of free self Ag reduced by off-target action of the Abs times  the duration of this reduction. Thus during the interval $[t_0,t_1]$ this damage is $A_S(\ab(t_0))\left(t_1-t_0\right)-\mathcal{D}_S\left(t_1-t_0\right)$.

\subsection*{Determination of optimal antibody dynamics}

We minimise the action [\ref{def:S}] subject to the constraint [\ref{def:self-damage}], i.e. we assume that removal of some amount of self Ag can be tolerated. This is equivalent~\cite{Gelfand2000} to minimisation of  [\ref{def:S}] with the  Lagrangian
\begin{eqnarray}
 \Lagr\left(\ab, \frac{\rmd}{\rmd t}{\ab}  \right)=  A_T\left(\ab\right)-\mathcal{T}\left(\frac{\rmd}{\rmd t}{\ab}\right)-\gamma A_S(\ab)  \label{eq:Lagrangian},
\end{eqnarray}
where $\gamma$ is a Lagrange parameter. The solution of the minimization  is described by   the Euler-Lagrange equation (see \MM):
\begin{eqnarray}
\frac{\rmd}{\rmd t}\frac{\partial}{\partial (\rmd{b}_\mu/\rmd t)} \mathcal{T}\left(\frac{\rmd}{\rmd t}{\ab}\right)=-\frac{\partial}{\partial b_\mu}   \left[A_T(\ab) - \gamma A_S(\ab) \right]\label{eq:E-L}.
%
%
\end{eqnarray}
We note that the above second order differential equations that describe the optimal control of antibody release were derived from general system level principles, with only minimal and plausible assumptions. 
Their solution will involve   $2M$ constants,  fixed by the boundary conditions $\ab(t_0)$ and  $\ab(t_1)$. 

The natural form for the kinetic energy is $\mathcal{T}(\rmd{\ab}/\rmd t)=\frac{1}{2}\sum_{\mu=1}^M\Lambda_\mu (\rmd b_\mu/\rmd t)^2$, where $\Lambda_\mu>0$. It  corresponds to assuming that at least one set of further (as yet unspecified) variables play a role in the Ab delivery process.  Insertion into [\ref{eq:E-L}]  gives us the  `Newtonian' equation
\begin{eqnarray}
\Lambda_\mu\frac{\rmd^2}{\rmd t^2}{b}_\mu=
A_T(\ab)\,r_\mu\!\left(\ab\right) - \gamma A_S(\ab)\,r^S_\mu\!\left(\ab\right),\label{eq:E-L-quadr}
\end{eqnarray}
where  we  used the affinities [\ref{def:affinity}] to express  the partial derivatives in  [\ref{eq:E-L}].  We note that the $\Lambda_\mu$, which reflect  properties of the Ab delivery mechanism,  act to introduce `inertia':  large (small)  $\Lambda_\mu$ reduce (increases) the tendency to change $\rmd{b}_\mu/\rmd t$.  
The  total `force' $\Lambda_\mu (\rmd^2{b}_\mu/\rmd t^2)$ in [\ref{eq:E-L-quadr}] is a sum of a target Ag dependent term  $A_T(\ab)\,r_\mu\!\left(\ab\right)$ that increases the rate of Ab delivery, and a self Ag dependent term  $- \gamma A_S(\ab)\,r^S_\mu\!\left(\ab\right)$ which decreases Ab delivery (if  $\gamma>0$).  The state of mechanical {equilibrium}  $\Lambda_\mu(\rmd^2{b}_\mu/\rmd t^2)=0$, marking the  balance of forces in  [\ref{eq:E-L-quadr}],  gives us, for $A_S(\ab), r_\mu\!\left(\ab\right)>0$, the identity 
\begin{eqnarray}
\frac{A_T(\ab)}{ \gamma A_S(\ab)}&=&\frac{r^S_\mu\!\left(\ab\right)}{r_\mu\!\left(\ab\right)}\label{eq:equilibrium}.
\end{eqnarray}
It follows that  there exists a function $\alpha(\ab)$ such that $r_\mu\!\left(\ab\right)=\alpha(\ab)\,r^S_\mu\!\left(\ab\right)$ for all $\mu$.  Furthermore,  for $\ab=\nullv$ the latter gives us the relation $r_\mu =\alpha \,r^S_\mu$ between affinities, where $\alpha=\alpha(\nullv)$.

  \section*{Results\label{section:results}}

  \subsection*{Free Ag  reduced by large numbers of `weak'  antibodies} \label{ssection:chemical-kinetics}
  To proceed with our model we need to determine the dependencies of $A_T$ and $A_S$ on the antibody amounts $\ab=(b_1,\ldots,b_M)$.
Here we consider   $M$ distinct  univalent  Abs  ${\I}_\mu$,  labelled by $\mu=1,\ldots,M$, each interacting with the univalent  target Ag ($\triangle$) and  self-Ag ($\circ$), via the following  chemical reactions  
  %
   \begin{eqnarray}
\circ +{\I}_\mu   \overset{    {K_{\mu}^{S+}}   }{ \underset{    {K_{\mu}^{S-}}   }{\rightleftharpoons} }  \overset{\circ}{\I}_\mu 
\label{eq:multi-Ab-Ag-reactions}
~~~~~~~~
 %
 %
\triangle +{\I}_\mu  \overset{  K^+_{\mu }    }{ \underset{  K^-_{\mu}      }{\rightleftharpoons} }  \overset{\triangle }{\I}_\mu
%
%
\end{eqnarray}
In chemical equilibrium, given the initial  concentrations $A_T(\nullv)$ of the target Ag and $A_S(\nullv)$ of the self-Ag,  the concentrations $A_T(\ab)$ of free target Ag and $A_S(\ab)$  of self-Ag are obtained by solving the following recursive system of equations; see \emph{Supplementary Information} (\SI), Section 1A:
  \begin{eqnarray}
A_T&=&\frac{A_T(\nullv)}{1 +\sum_{\mu=1}^Mb_\mu \frac{r_\mu     }{1+A_T\,r_\mu+A_S\,r^S_\mu}}\label{eq:Ag-recursion} \\
A_S&=&\frac{A_S(\nullv)}{1 +\sum_{\mu=1}^M b_\mu\frac{r^S_\mu       }{1+A_T\,r_\mu+A_S\,r^S_\mu}}.
\end{eqnarray}
Each Ab is characterised by its affinities to the target Ag, $r_\mu=K_{\mu}^{+}/K_{\mu}^{-}$ (the ratio of forward and backward rates), and self-Ag, $r^S_\mu=K_{\mu}^{S+}/K_{\mu}^{S-}$. These  give rise to the affinity vectors $\rv=(r_1,\ldots,r_M)$ and $\rv^S=(r_1^S,\ldots,r_M^S)$,  which define  the  Ab \emph{repertoire}.  For multiple self-Ags the repertoire is a matrix of affinities (see \SI, Sections 1A \& 2A).

In order to use [\ref{eq:E-L-quadr}] one would prefer an explicit expression for   $A_T\!\left(\ab\right)$ and $A_S\!\left(\ab\right)$,  but  how to solve  the non-linear recursion  [\ref{eq:Ag-recursion}] analytically is not clear.  However, if we assume that  affinities scale as $r_{\mu}\equiv r_{\mu}/M$ and  $r_{\mu}^S\equiv r_{\mu}^S/M$, then in the regime $M\rightarrow\infty$ of having a large number of individually weak Abs,  we obtain the concentrations of free  Ags in explicit form (see \MM):
  \begin{eqnarray}
A_T\left(\ab\right)=\frac{A_T(\nullv)}{1 +  B(\ab)        },~~~~~~~~\label{eq:Ag-MF}
A_S\left(\ab\right)=\frac{A_S(\nullv)}{1 +  B_S(\ab)       }
\end{eqnarray}
expressed as  functions  of the averages
  \begin{eqnarray}
B_T(\ab)=\frac{1}{M}\sum_{\mu=1}^Mr_\mu\, b_\mu   \label{def:B},~~~~~~
B_S(\ab)=\frac{1}{M}\sum_{\mu=1}^M   r^S_\mu\, b_\mu      \nonumber.
\end{eqnarray}
The averages  $B_T(\ab)$ and $B_S(\ab)$  can be seen as \emph{total affinities}  to the target Ag and the self Ag.  A similar object, where $b_\mu$ was the  number of B cells with affinity to Ag $r_\mu/M$,  was postulated as an  `energy' function of somatic hypermutation in~\cite{Kepler1993}.
 
 We note that  the result [\ref{eq:Ag-MF}],  although  derived for univalent Abs and Ag,  is also true for multivalent  Abs (see \SI, Section 1B). Thus our model predicts that it is possible to reduce target antigen without requiring affinity-matured antibodies, such as those produced in a T-dependent reaction, if a sufficient number of weaker binders are available.  Furthermore, the framework outlined here can easily incorporate   multiple Ags, chemical species binding Ab-Ag complexes,  phagocytes, etc. (see \emph{SI}, Section 1A)
   
\subsection*{Reduced macroscopic description} \label{ssection:Ab-dynamics}

Let us consider the Euler-Lagrange equations [\ref{eq:E-L-quadr}] for  the free and self-Ag.  Via [\ref{eq:Ag-MF}], and upon reverting from the right-hand side of [\ref{eq:E-L-quadr}] back to that of  its predecessor [\ref{eq:E-L}], these  now take the form 
\begin{eqnarray}
\Lambda_\mu\frac{\rmd^2}{\rmd t^2}{b}_\mu
=\frac{A_T(\nullv) }{\left(1+ B_T\right)^2 }\frac{ r_\mu}{M} - \gamma \frac{A_S(\nullv)}{ \left(1+   B_S \right)^2  } \frac{ r^S_\mu}{M}    \label{eq:E-L-MF},
\end{eqnarray}
where  $B_T\equiv  B(\ab)$ and $B_S\equiv  B_S(\ab) $.  If we assume that $\Lambda_\mu$ scales as  $\Lambda_\mu= \lambda_\mu\phi(M) /M$,  where  $\phi(M)= o(M)$, we can derive for $M\to\infty$ the following equations   (\emph{SI}, Section 2A):
\begin{eqnarray}
\frac{\rmd^2}{\rmd t^2}{B}_T&=&  \frac{A_0^T     \vert \rv\vert^2}{\left(1+  B_T \right)^2 } - \gamma \frac{A_0^S    (\rv\cdot\rv^S)}{ \left(1+     B_S \right)^2  }    \label{eq:MF-dynamics}\\
\frac{\rmd^2}{\rmd t^2}{B}_S&=&\frac{A_0^T  (\rv\cdot\rv^S)   }{\left(1+  B_T \right)^2 } - \gamma \frac{A_0^S    \vert \rv^S \vert^2}{ \left(1+    B_S \right)^2  }     \nonumber,
\end{eqnarray}
where in the above we used the dot product definition $\xv \cdot\yv= M^{-1}\sum_{\mu=1}^M \lambda_\mu^{-1}    x_{\mu}   y_{\mu}  $,  with the associated norm  $\vert \xv\vert=\sqrt{\xv \cdot\xv}$.  We assume that at time $t=0$ all Ab amounts and production rates are zero, i.e.  $b_\mu=\rmd{b}_\mu/\rmd t=0$ for all $\mu$, so the initial conditions for [\ref{eq:MF-dynamics}] are   $B_T(0)=B_S(0)=0$ and $(\rmd{B}_T/\rmd t)(0)=(\rmd{B}_S/\rmd t)(0)=0$.   Furthermore, the average Ab concentration $\tilde{B}(t)=M^{-1}\sum_{\nu=1}^M b_\nu(t)$  is governed by the equation 
\begin{eqnarray}
\frac{\rmd^2}{\rmd t^2}{\tilde{B}}&=&  \frac{A_0^T   ( \rv\cdot\1)  }{\left(1+  B_T \right)^2 } - \gamma \frac{A_0^S    ( \rv^S\!\cdot\1)}{ \left(1+     B_S \right)^2  }    \label{eq:MF-Ab-dynamics}.
\end{eqnarray}
with the short-hand  $\1=(1,\ldots,1)$. 

The simplest case to consider is that where each Ab is either self-reactive or non-self-reactive, i.e. for each $\mu$ either $r_\mu>0$ or  $r^S_\mu>0$, but never both.  This  implies  that $\rv\cdot\rv^S=0$, and that hence  [\ref{eq:MF-dynamics}] decouples into two  independent equations:  
\begin{eqnarray}
\frac{\rmd^2}{\rmd t^2}{B}_T=  \frac{A_0^T    \vert \rv\vert^2    }{\left(1+  B_T \right)^2 },~~~~~~
\frac{\rmd^2}{\rmd t^2}{B}_S=- \gamma \frac{A_0^S \vert \rv^S \vert^2  }{ \left(1+    B_S \right)^2  }    
 \label{eq:MF-dynamics-Ab-dichotomy}
\end{eqnarray}
The dynamics of $B_T$  is now \emph{conservative},  with energy function
\begin{eqnarray}
E\Big(B_T, \frac{\rmd B_T}{\rmd t}\Big)=   \frac{1}{2    \vert \rv\vert^2}\Big(\frac{\rmd B_T}{\rmd t}\Big)^2 +  \frac{A_0^T }{1+  B_T} \label{def:E}, 
\end{eqnarray}
where the terms $(\rmd B_T/\rmd t)^2/2 \vert \rv\vert^2$ and $A_0^T /(1+  B_T)$ are, respectively, the `kinetic'  and `potential'  energies.   The equation for $B_T$   describes  the motion  of a  `particle'  of  `mass'  $ 1/\vert \rv\vert^2$ in in a potential field~\cite{Arnold1989}.  Furthermore, solving the energy conservation equation $E(B_T,\rmd B_T/\rmd t)=E(B_T(0), (\rmd B_T/\rmd t)(0))$, for  $B(0)=(\rmd B_T/\rmd t)(0)=0$, gives us
\begin{eqnarray}
\frac{\rmd}{\rmd t}{B_T}&=&\sqrt{2 A_0^T   \vert \rv\vert^2   \frac{    B_T   }{\left(1+  B_T \right) } } \label{eq:MF-dB/dt}. 
\end{eqnarray}
The function $\sqrt{B_T/(1+  B_T) }\in[0,1]$ is monotonic increasing and concave for $B_T\geq 0$.  Hence 
t $B(t)$ is bounded from above by $\sqrt{2 A_0^T \vert \rv\vert^2   }\, t $ and  this bound is saturated as  $t\rightarrow\infty$. Also, the (normalised) amount    of target antigen  $A_T(\ab(t))/A_T(\nullv)=(1+  B_T(t))^{-1}$ is bounded from below by $(1+  \sqrt{2 A_0^T\vert \rv\vert^2   }  \, t  )^{-1}$.  

In a similar manner we simplify the dynamics of $B_S$, which   is also conservative, describing the motion of a particle of mass $\vert \rv^S\vert^{-2}$ and potential energy $-\gamma   A_0^S /(1+  B_S)$. Here we find
\begin{eqnarray}
\Big(\frac{\rmd}{\rmd t}{B}_S\Big)^2=-2\gamma A_0^S  \vert \rv^S\vert^2  \frac{  B_S    }{\left(1+  B_S \right) }  \label{eq:MF-dBs/dt-1}
\end{eqnarray}
Since  $\gamma>0$  and with the assumed initial conditions,  the (trivial) solution is $B_S=0$, i.e. self-reactive Abs are \emph{not} used. 

We have now seen that   [\ref{eq:MF-dynamics-Ab-dichotomy}]   can be mapped into equations  of Classical Mechanics.  The equation for $B_S$ describes the acceleration of a particle of  mass $\vert \rv^S\vert^{-2}$ in a  gravitational  field with gravitational constant $\gamma  $,  created by a another particle of mass $A_0^S$ and radius one~\cite{Arnold1989}.  The equation for $B_T$ has a  similar interpretation  but with a {repulsive}  potential.

\subsection*{Ag removal is faster in a more diverse repertoire, and slower when the repertoire has higher self-reactivity}\label{ssection:steady-state}%

We return to the more general case where $\rv\cdot\rv^S>0$, so Abs may have the potential to bind both target Ag and self Ag.
Further analytic results can be obtained in the equilibrium regime of  [\ref{eq:MF-dynamics}], defined by $\rmd^2 B_T/\rmd t^2=\rmd^2 B_S/\rmd t^2=0$. This can only occur when   $r_\mu=\alpha r^S_\mu$ for all $\mu$ (see {\em SI}, Section 2B) , where $\alpha>0$. The inverse $\alpha^{-1}$ can be seen as  a \emph{degree of self-reactivity}.  From [\ref{def:B}] it follows that  $B_T=\alpha B_S$ in this regime,  and that   [\ref{eq:MF-dynamics}] can be reduced to a single  equation:
\begin{eqnarray}
\frac{\rmd^2}{\rmd t^2}{B}_S=A_0^S\vert \rv^S\vert^2   \left[\frac{\alpha\beta  }{\left(1+  \alpha B_S \right)^2 } - \frac{ \gamma}{ \left(1+    B_S \right)^2  }  \right]    \label{eq:MF-Bs-ode},
\end{eqnarray}
with $\beta =A_0^T/A_0^S$.  It is easy to show, using the above  equation and   [\ref{eq:MF-Ab-dynamics}], that now $\rmd^2\tilde{B}/\rmd t^2=(\rv^S\!\cdot\1)    \vert \rv^S\vert^{-2} \rmd^2{B}_S/\rmd t^2$,  and hence the average concentration of Abs is given by 
\begin{eqnarray}
\tilde{B}=   (\rv^S\!\cdot\1)    \vert \rv^S\vert^{-2}    B_S
\label{eq:MF-Ab-equality}.
\end{eqnarray}
The dynamics  [\ref{eq:MF-Bs-ode}] is again conservative, now with energy 
\begin{eqnarray}
E\Big(B_S, \frac{\rmd {B}_S}{\rmd t}\Big)=\frac{1}{2\vert \rv^S\vert^2}\Big(\frac{\rmd {B_S}}{\rmd t}\Big)^2+  \frac{\beta  A_0^S}{1+  \alpha B_S } - \frac{ \gamma A_0^S}{1+    B_S}.       \label{def:E-self}
\end{eqnarray}
As before we can use energy conservation, following initial conditions $B_S(0)=(\rmd B_S/\rmd t)(0)=0$, 
to derive 
\begin{eqnarray}
\frac{\rmd}{\rmd t}{B}_S=\sqrt{\!2A_0^S \vert \rv^S\vert^2 \! \left   (
\frac{\beta\alpha B_S}{1+\alpha B_S}-\frac{\gamma B_S}{1+B_S}
 \right)} \label{eq:MF-velocity}. 
\end{eqnarray}
From this follows the following upper bound,  which is saturated as $t\to\infty$
(see {\em SI}, Section 2B):
\begin{eqnarray}
B_S(t)\leq t/\tau,\label{eq:MF-B-ub}
\end{eqnarray}
with the time constant  
\begin{eqnarray}
\tau=1/ \vert \rv^S\vert\sqrt{2A_0^S(\beta-\gamma)}.
\label{def:MF-time-const}
\end{eqnarray} 
 As a  consequence of  [\ref{eq:MF-B-ub}], we find for  the normalised target Ag  
\begin{eqnarray}
\frac{A_T(\ab(t))}{A_T(\nullv)}=\frac{1}{1+\alpha B_S(t)}\geq \frac{1}{1+\alpha t/\tau}\label{eq:MF-Ag-lb}.
\end{eqnarray}
So $\tau/\alpha$ is a lower bound for the \emph{half-life} of free target Ag;  to achieve  $A_T(\ab(t))/A_T(\nullv)=\frac{1}{2}$, the required time $t$ has to be at least  $\tau/\alpha$. The lower bound for the  half-life of self-Ag, derived by a similar argument, is found to be $\tau$.  Furthermore,  if we define $w(\lambdav)=M^{-1}\sum_{\mu=1}^M\lambda^{-1}_\mu$ then
\begin{eqnarray}
 \vert \rv^S\vert = \sqrt{w(\lambdav)\left[ \sigma_{\lambdav}^2\left(\rv^S\right)+  m_{\lambdav}^2\left( \rv^S \right) \right]  } \label{eq:MF-2-nd-moment},
\end{eqnarray}
where $\sigma_{\lambdav}^2\left(\rv^S\right)=\vert \rv^S\vert^2/w(\lambdav)   -  ((\rv^S\!\cdot\1)/ w(\lambdav))^2$
and $m_{\lambdav}( \rv^S)= (\rv^S\!\cdot\1)/ w(\lambdav)  $ are, respectively,  variance and mean of the self-affinities $\rv^S$ (see {\em SI}, Section 2B).  
%
%
Thus  $\tau$ is monotonically decreasing with the variance $\sigma_{\lambdav}^2(\rv^S)$ and the mean  $m_{\lambdav}( \rv^S)$.  Since the former can be seen as a measure of the repertoire's `diversity',  having a more diverse repertoire facilitates  a more rapid  reduction of target Ag. 
 
We also solved the differential  equation [\ref{eq:MF-Bs-ode}] numerically  for different inverse self-reactivities $\alpha$.   The solutions are plotted in {\em Supplementary Information},   in Figures 5-8.  
 Comparison of  the upper bound  [\ref{eq:MF-B-ub}] with the solutions of  [\ref{eq:MF-Bs-ode}]  in Figure 9  allows us to summarise  various regimes.  We first define, using  [\ref{def:damage}], the normalised damage per unit time  $\delta_A(t_1-t_0)=\mathcal{D}_A\left(t_1-t_0\right)/A_T(\ab(t_0))\left(t_1-t_0\right)$, 
 %
%
where $0\leq\delta_A\leq1$, and, using   [\ref{def:self-damage}], the normalised damage to self per unit time $1-\delta_S(t_1-t_0)=1-\mathcal{D}_S\left(t_1-t_0\right)/A_S(\ab(t_0))\left(t_1-t_0\right)$, 
 %
%
where $0\leq\delta_S\leq1$  and $0\leq1-\delta_S\leq1$.  For the system [\ref{eq:Ag-MF}],  on the time interval $[0,t]$, the above definitions give us 
 \begin{eqnarray}
\delta_A(t)=\frac{1}{t}   \int_{0}^{t}   \!     \frac{\rmd t^\prime}{1+  \alpha B_S({t^\prime})},~~~~
\delta_S(t)=\frac{1}{t} \int_{0}^{t}  \!   \frac{\rmd t^\prime}{1+  \!B_S({t^\prime})} 
\label{eq:MF-damage}
\end{eqnarray}
Now since $(1+  \alpha B_S)^{-1}$ is a monotonic decreasing function of $B_S$, the upper bound   [\ref{eq:MF-B-ub}] gives us the lower bounds
 \begin{eqnarray}
\delta_A(t)&\geq &\frac{\tau}{\alpha \, t}\log\left(1+  \frac{\alpha \,t}{\tau}\right)\label{eq:MF-damage-lb}
\\
\delta_S(t)  &\geq&\frac{\tau}{t}\log\left(1+  \frac{t}{\tau} \right)   \label{eq:MF-self-damage-lb}.
\end{eqnarray}
The latter gives us the upper  bound $1-(\tau/t)\log(1+ t/\tau)\geq 1-\delta_S(t)$ for the  damage to self.  

\begin{figure}
\centering
\includegraphics[width=.8\linewidth]{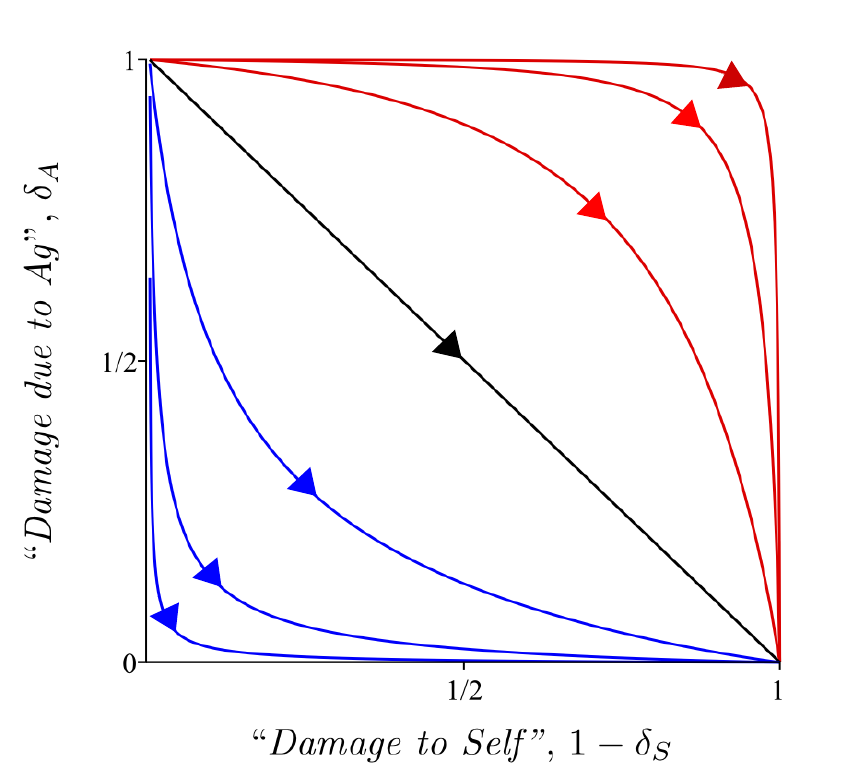}
\caption{The  damage due to antigen $\delta_A$ (lower bound), plotted   as a function of  the  damage to self $1-\delta_S$ (upper bound) for the  inverse self-reactivity $\alpha=\{10^{-3}, 10^{-2}, 10^{-1}\}$ (top red curves with $\alpha$ increasing from top to bottom) , $\alpha=1$ (black line) and   $\alpha=\{10, 10^2, 10^3\}$ (bottom blue curves with $\alpha$ increasing from top to bottom). The direction of `time' $t/\tau\in[0,\infty)$, indicated by arrows, is always from left to right. }
 \label{figure:MF-damage} 
\end{figure}

The two bounds on damages are plotted in  Figure  \ref{figure:MF-damage} for different values of self-reactivity constant $\alpha$.  For  a repertoire with Abs binding $\alpha$ times stronger to the target Ag  than to the self-Ag the immune response is  `normal'   and `autoimmune', respectively,  when $\alpha>1$ and $\alpha<1$.  The normal response is characterised by a  large  decrease of free target Ag and  a small decrease in free self-Ag per unit of time. For the autoimmune response it is the  opposite.  Furthermore,   the  normal response  is `accelerated'  by   a  larger $\alpha$ and increased repertoire diversity,  but,  for the same repertoire diversity,  the autoimmune response is slower.

\section*{Discussion \label{section:discussion}}
In this work we have shown, using only minimal assumptions, that antibody repertoire diversity is important in the effective removal of antigen, in multiple ways.  Not just because the repertoire will then have more chance of containing a single dominant antibody that can react to the target-Ag, but also because for a more  diverse repertoire  the half life of target-Ag will be smaller.  Hence any decrease in repertoire diversity, such as that observed in older age, or caused by a prior immune response, can have an adverse effect on the immune response to challenge.  Furthermore, reduction in efficacy of central tolerance mechanisms such as can occur in older age, will result in greater self-reactivity in the repertoire, and this too will hamper an efficient immune response against target-Ag.

The mathematical framework  in the form  developed here can for now only be used  to model the immune response  to a finite amount of Ag, with a fixed repertoire of Abs. Adaptation of the affinities of Abs to target Ag via  affinity maturation~\cite{Janeway2012} is not yet included.   To model the latter on could modify the Lagrangian  [\ref{eq:Lagrangian}],  and derive  dynamic equations for  affinities.   Also the present restriction on the amount of Ag can be relaxed within the current framework, by introducing (partially stochastic)  Ag reproduction and death.

\matmethods{
\subsubsection*{The Variational Problem}
We aim to find the path $\ab(t)$ that minimises the action [\ref{def:S}] on the time-interval $[t_0, t_1]$ with the boundaries $\ab(t_0)=\ab_0$ and $\ab(t_1)=\ab_1$.  This path  must solve the equation $\delta\mathcal{S}=0$  for the  difference $\delta\mathcal{S}=\mathcal{S}(\ab+\delta\ab, \rmd{\ab}/\rmd t+\rmd\delta{\ab}/\rmd t)-\mathcal{S}(\ab, \rmd{\ab}/\rmd t)$,  where  $\ab(t)+\delta\ab(t)$  is  any \emph{perturbed} path with $\delta\ab(t_0)=\delta\ab(t_1)=0$~\cite{Gelfand2000}.  Using  the differential operator $ \nabla_{\ab}=\left(\partial/\partial b_1, \ldots, \partial/\partial b_M\right)$ this difference, up to the order $O\!\left({\vert\delta\ab\vert}^2\right)$, can be written in the form 
%
\begin{eqnarray}
\delta\mathcal{S}&=&\int_{t_0}^{t_1} \Lagr\left(\ab\!+\!\delta\ab, \frac{\rmd{\ab}}{\rmd t}\!+\!\delta\frac{\rmd{\ab}}{\rmd t}\right)\,\rmd t-\!\int_{t_0}^{t_1}\Lagr\left(\ab, \frac{\rmd{\ab}  }{\rmd t}\right)\,\rmd t\label{eq:dS}\\
&=&\int_{t_0}^{t_1}\!\!\!\!  \left\{\delta\ab. \nabla_{\ab} \Lagr\left(\ab, \frac{\rmd{\ab}}{\rmd t}\right) + \delta\frac{\rmd{\ab}}{\rmd t}. \nabla_{\rmd\ab/\rmd t}\Lagr\left(\ab, \frac{\rmd{\ab}}{\rmd t}\right) \right\}\!\rmd t 
\nonumber\\
&=&\left[\delta\ab\,. \nabla_{\rmd\ab/\rmd t}\Lagr\left(\ab, \frac{\rmd{\ab}}{\rmd t}\right)  \right]_{t_0}^{t_1}+\cdots \nonumber\\
&&\cdots+ \int_{t_0}^{t_1} \delta\ab\,. \Big\{ \nabla_{\ab}\Lagr\left(\ab, \frac{\rmd{\ab}}{\rmd t}\right)-\frac{\rmd}{\rmd t} \nabla_{\rmd\ab/\rmd t}\Lagr\left(\ab,\frac{\rmd{\ab}}{\rmd t}\right) \Big\}\rmd t\nonumber\\
&=&\int_{t_0}^{t_1}\delta\ab\,. \!\left\{ \nabla_{\ab}\Lagr\left(\ab, \frac{\rmd{\ab}}{\rmd t}\right) -\frac{\rmd}{\rmd t} \nabla_{\rmd\ab/\rmd t}\Lagr\left(\ab, \frac{\rmd{\ab}}{\rmd t}\right)  \right\}\rmd t \nonumber
\end{eqnarray}
where we used integration  by parts and the stated boundary conditions.  Solving $\delta\mathcal{S}=0$ for the part of $\delta\mathcal{S}$ that is linear in $\delta\ab$ gives us the so-called Euler-Lagrange equation 
\begin{eqnarray}
\frac{\rmd}{\rmd t} \nabla_{\left(\rmd{\ab}/\rmd t\right)}\Lagr\left(\ab, \frac{\rmd{\ab}}{\rmd t}\right)= \nabla_{\ab}\Lagr\left(\ab, \frac{\rmd{\ab}}{\rmd t}\right)   \label{eq:Euler-Lagrange}
\end{eqnarray}
with boundary conditions $\ab(t_0)=\ab_0$ and $\ab(t_1)=\ab_1$. 

\subsubsection*{Mean-Field Limit}

Here we explain briefly the derivation of [\ref{eq:Ag-MF}] from [\ref{eq:Ag-recursion}]. 
Substituting  $r_\mu\to r_\mu/M$ and $r^S_\mu\to r^S_\mu/M$ into [\ref{eq:Ag-recursion}] gives
  \begin{eqnarray}
\frac{A_T}{A_T(\nullv)}=\frac{1}{1 +\frac{1}{M}\sum_{\mu=1}^M r_\mu b_\mu (1+A_T\,r_\mu/M+A_S\,r^S_\mu/M)^{-1}}\label{eq:Ag-recursion-A_again}
\end{eqnarray}
hence, if  $A_T(\nullv)=\phi(M)A_0^T$ and  $A_S(\nullv)=\phi(M)A_0^S$,  where $\phi(M)= o(M)$, i.e. $\lim_{M\rightarrow\infty}  \phi(M)/M=0$,  then for $M\to\infty$ we will indeed find  the mean-field expressiom [\ref{eq:Ag-MF}] since
  \begin{eqnarray}
\frac{A_T}{A_T(\nullv)}=\frac{1}{1 +\frac{1}{M}\sum_{\mu=1}^M r_\mu b_\mu +O(\phi(M)/M)} 
\end{eqnarray}
}

\showmatmethods{} 

\acknow{This work was supported by the Medical Research Council of the  United Kingdom (grant MR/L01257X/1).  A.M. would like to thank  Adriano Barra, Alessia Annibale,  Fabi\'{a}n Aguirre L\'{o}pez, Attila Csik\'{a}sz-Nagy and Mart\'{i}  Aldea Malo for very enlightening discussions.}

\showacknow{} 

\section*{Supplementary Information}

\section{Chemical kinetics of antigen-antibody reactions\label{section:chemical-kinetics}}
 
  \subsection{Univalent antibodies  reacting with univalent antigens\label{ssection:univalent-Ab-Ag-reactions} }
We consider   $M$ different  univalent  antibodies (Abs), represented by the symbols ${\I}_\mu$ with $\mu\in\{1,\ldots,M\}$, forming complexes  with $M_A$ different univalent target antigens (Ags), $\triangle_v$ with $v\in\{1,\ldots,M_A\}$,  and $M_S$ self-Ags, $\circ_u$ with  $u\in\{1,\ldots,M_S\}$.  The Ag bound by Ab $\overset{\triangle_v }{\I_\mu}$ and $\overset{\circ_u}{\I_\mu}$ will subsequently form complexes with `phagocytic'  species $\Ph$~\cite{Janeway2012}.  The formation and dissociation of complexes is modelled by the four chemical reactions   
   \begin{eqnarray}
\circ_u +{\I}_\mu   \overset{    {K_{\mu u}^{S+}}   }{ \underset{    {K_{\mu u}^{S-}}   }{\rightleftharpoons} }~  \overset{\circ_u}{\I_\mu} ~~~~~~~~~~
 \overset{\circ_u }{\I_\mu} +\Ph  \overset{     K^+    }{ \underset{    K^-   }{\rightleftharpoons} }    ~ \overset{\circ_u }{\I_\mu}\Ph 
 ~~~~~~~~~~
\triangle_v +{\I}_\mu  \overset{  K^+_{\mu v}    }{ \underset{  K^-_{\mu v}      }{\rightleftharpoons} }~  \overset{\triangle_v }{\I_\mu}~~~~~~~~~~
\overset{\triangle_v }{\I_\mu} +\Ph  \overset{    K^+  }{ \underset{    K^-   }{\rightleftharpoons} } ~  \overset{\triangle_v }{\I_\mu}\Ph.
\label{eq:multi-Ab-Ag-reactions-2}
\end{eqnarray}
In \emph{chemical equilibrium}~\cite{Yablonskii1991} the concentrations of \emph{free} self-Ag,  target Ag, Ab and P (denoted, respectively, by the symbols $\left[\circ_u\right]$, $\left[\triangle_v\right]$, $\left[\I_\mu\right]$ and $\left[\Ph\right]$) are related to the concentration of \emph{bound} species $\overset{\circ_u }{\I_\mu}$, $\overset{\circ_u }{\I_\mu}\Ph$, $\overset{\triangle_v }{\I_\mu}$ and  $\overset{\triangle_v }{\I_\mu}\Ph$ (denoted, respectively, by the symbols  $[\overset{\circ_u }{\I_\mu}  ]$, $[\overset{\circ_u }{\I_\mu}\Ph]$, $[\overset{\triangle_v }{\I_\mu}  ]$ and $[\overset{\triangle_v }{\I_\mu}\Ph]$)  via the \emph{affinity} parameters  $r_{\mu u}^S={K_{\mu u}^{S+}} /{K_{\mu u}^{S-}}$, $r_{\mu v}={K^+_{\mu v}} {K^-_{\mu v}}$ and $r=K^+/K^-$, i.e. the ratios of forward/backward rates of reactions:
\begin{eqnarray}
r_{\mu u}^S =\frac{  [\overset{\circ_u }{\I_\mu}  ]     }{ \left[\circ_u\right]     \left[\I_\mu\right] } ~~~~~~~~
r=   \frac{  [ {\overset{\circ_u }{\I_\mu}} \Ph  ]     }{  [\overset{\circ_u }{\I_\mu}  ]      \left[\Ph\right] } ~~~~~~~~
r_{\mu v} =   \frac{  [\overset{\triangle_v }{\I_\mu} ]     }{ \left[\triangle_v\right]     \left[\I_\mu\right] } ~~~~~~~~
r=   \frac{  [\overset{\triangle_v }{\I_\mu} \Ph]     }{  [\overset{\triangle_v }{\I_\mu} ]    \left[\Ph\right] }
\label{eq:multi-Ab-Ag-equilibrium}
\end{eqnarray}
Upon denoting the initial  concentrations of  the species $\circ_u$, $\triangle_v$, $\I_\mu$ and $\Ph$ by  $\left[\circ_u\right]_0$, $\left[\triangle_v\right]_0$, $\left[\I_\mu\right]_0$ and $\left[\Ph\right]_0$,  we can use \emph{mass conservation} to write
\\[-6.5mm]
\begin{eqnarray}
\left[\circ_u\right]_0&=&\left[\circ_u\right]+  \sum_{\mu=1}^M\,  [\overset{\circ_u }{\I_\mu}  ]  + \sum_{\mu=1}^M\,[\overset{\circ_u }{\I_\mu}\Ph  ]  \label{eq:multi-Ab-Ag-mass-1}\\
\left[\triangle_v\right]_0&=&\left[\triangle_v\right]+  \sum_{\mu=1}^M\,  [\overset{\triangle_v }{\I_\mu}  ]  +  \sum_{\mu=1}^M\,[\overset{\triangle_v }{\I_\mu} \Ph]  \\
\left[\I_\mu\right]_0&=&\left[\I_\mu\right]+  \sum_{u=1}^{M_S}\,  [\overset{\circ_u }{\I_\mu}  ]+  \sum_{v=1}^{M_A}\,  [\overset{\triangle_v }{\I_\mu}  ]  +  \sum_{u=1}^{M_S}\,  [\overset{\circ_u }{\I_\mu}  \Ph]+  \sum_{v=1}^{M_A}\,  [\overset{\triangle_v }{\I_\mu}  \Ph] \\
\left[\Ph\right]_0&=&\left[\Ph\right]+    \sum_{\mu=1}^M\sum_{u=1}^{M_S}\,  [\overset{\circ_u }{\I_\mu}  \Ph]+  \sum_{\mu=1}^M\sum_{v=1}^{M_A}\,  [\overset{\triangle_v }{\I_\mu}  \Ph] 
\end{eqnarray}
By using [\ref{eq:multi-Ab-Ag-equilibrium}] these expressions can be written in the alternative form
\\[-3.5mm]
\begin{eqnarray}
\left[\circ_u\right]_0&=&\left[\circ_u\right]\Big( 1+\left(1+r[\Ph]\right)\sum_{\mu=1}^M r_{\mu u}^S  [{\I}_\mu  ]\Big)   \label{eq:multi-Ab-Ag-mass-2}\\
\left[\triangle_v\right]_0&=&\left[\triangle_v\right]\Big(1+ \left(1+r[\Ph]\right) \sum_{\mu=1}^M  r_{\mu v} [{\I}_\mu  ]  \Big) \nonumber\\
\left[\I_\mu\right]_0&=&\left[\I_\mu\right]\Big(1+  \left(1+r[\Ph]\right) \left\{  \sum_{u=1}^{M_S} r_{\mu u}^S  [\circ_u ]+  \sum_{v=1}^{M_A} r_{\mu v} [\triangle_v   ] \right\}\Big) \nonumber\\
\left[\Ph\right]_0&=&\left[\Ph\right]\Big( 1    +    r\sum_{\mu=1}^M \left\{ \sum_{u=1}^{M_S} r_{\mu u}^S  [\circ_u ]     + \sum_{v=1}^{M_A} r_{\mu v} [\triangle_v   ] \right\} \left[\I_\mu\right]   \Big)\nonumber.
\end{eqnarray}
Finally, upon introducing the notation $A^S_u$ and $A^T_v$ for the concentrations $ \left[\circ_u\right]$ of free self Ags and $\left[\triangle_v\right]$ of target   Ags we obtain  the following system of  recursive equations, which, given the initial concentrations  $A^S_u(\nullv)\equiv\left[\circ_u\right]_0$ and $A^T_v(\nullv)\equiv\left[\triangle_v\right]_0$,  $b_\mu \equiv\left[\I_\mu\right]_0$ and  $P(\nullv)\equiv\left[\Ph\right]_0$, can be used to obtain   the equilibrium concentrations of free self and target Ag:
\\[-3.5mm]
\begin{eqnarray}
 A^S_u &=&   \frac{A^S_u(\nullv)}{  1+\left(1+r[\Ph]\right)\sum_{\mu=1}^M r_{\mu u}^S  [{\I}_\mu  ] }   \label{eq:multi-Ab-Ag-recursion-1},~~~~~~~~
A^T_v =             \frac{  A^T_v(\nullv) }{    1+ \left(1+r[\Ph]\right) \sum_{\mu=1}^M  r_{\mu v} [{\I}_\mu  ]   }\\
\left[\I_\mu\right] &=&      \frac{  b_\mu  }        { 1+  \left(1+r[\Ph]\right) \left\{  \sum_{u=1}^{M_S} r_{\mu u}^S A^S_u+  \sum_{v=1}^{M_A} r_{\mu v} A^T_v \right\}} \nonumber\\
 \left[\Ph\right]    &=&  \frac{   P(\nullv) }{  1    +    r\sum_{\mu=1}^M \left\{ \sum_{u=1}^{M_S} r_{\mu u}^S  A^S_u     + \sum_{v=1}^{M_A} r_{\mu v} A^T_v \right\} \left[\I_\mu\right]  }\nonumber
\end{eqnarray}
We  assume that the individual  antibody affinities are weak, i.e. $r_{\mu u}^S\equiv r_{\mu u}^S/M$ and  $r_{\mu v}\equiv r_{\mu v}/M$, and consider
\begin{eqnarray}
r \left[\Ph\right]    &=&  \frac{  r P(\nullv) }{   1    +    \frac{r}{M}\sum_{\mu=1}^M \left\{ \sum_{u=1}^{M_S} r_{\mu u}^S  A^S_u     + \sum_{v=1}^{M_A} r_{\mu v} A^T_v \right\} \left[\I_\mu\right]   }\label{eq:multi-Ab-Ag-MF-r[P]-1}\\
&=&  \frac{  r P(\nullv) }{   1    +    \frac{r}{M}\sum_{\mu=1}^M \left\{ \sum_{u=1}^{M_S} r_{\mu u}^S  \tilde{A}^S_u A^S_u(\nullv)    + \sum_{v=1}^{M_A} r_{\mu v}  \tilde{A}^T_v A^T_v(\nullv)\right\} \left[\I_\mu\right]   }\nonumber,
\end{eqnarray}
where we have defined the normalised  concentrations $\tilde{A}^T_v=A^T_v/A^T_v(\nullv) $ and  $\tilde{A}^S_u=A^S_u/  A^S_u(\nullv) $, in the limit $M\rightarrow\infty$ of a `large' number of Ab types.  If $M_A$ and $M_S$ are finite and  $A^S_u(\nullv), A^T_v(\nullv), P(\nullv) \propto\phi(M)$,  where we allow for $\phi(M)\rightarrow\infty$ as $M\rightarrow\infty$, but such that  $\phi(M)/M\rightarrow0$, i.e. $\phi(M)\in o(M)$,  then 
\begin{eqnarray}
r \left[\Ph\right]    &=&  \frac{  r P_0 }{    \frac{r}{M}\sum_{\mu=1}^M \left\{ \sum_{u=1}^{M_S} r_{\mu u}^S  \tilde{A}^S_u A_u^{0S}    + \sum_{v=1}^{M_A} r_{\mu v}  \tilde{A}^T_v A_v^{0T}\right\} \left[\I_\mu\right]  +  \frac{1}{\phi(M)}   }\label{eq:multi-Ab-Ag-MF-r[P]-2},
%
%
\end{eqnarray}
where $P(\nullv)=\phi(M)P_0$, $A^S_u(\nullv)=\phi(M)  A_u^{0S}$ and $A^T_v(\nullv)=\phi(M)  A^{0T}_v$. Thus $r \left[\Ph\right] = O(M^0)$ when  $r_{\mu u}^S, r_{\mu v}= O(M^{-1})$, $M_A, M_S= O(M^0)$ and $A^S_u(\nullv), A^T_v(\nullv), P(\nullv) = o(M)$.
 Using the above result in our equation for $[{\rm I}_\mu]$ gives
\begin{eqnarray}
\left[\I_\mu\right] &=&      \frac{  b_\mu  }        { 1+  \left(1+r[\Ph]\right) \left\{  \sum_{u=1}^{M_S} r_{\mu u}^S A^S_u+  \sum_{v=1}^{M_A} r_{\mu v} A^T_v \right\}} \label{eq:multi-Ab-Ag-MF-I}\\
&=&   \frac{  b_\mu  }        { 1+  \left(1+r[\Ph]\right)     \left\{  \sum_{u=1}^{M_S} r_{\mu u}^S  \tilde{A}^S_u A_u^{0S}+  \sum_{v=1}^{M_A} r_{\mu v}   \tilde{A}^T_v A_v^{0T}      \right\} \frac{\phi(M)}{M}  }  \nonumber \\
&=&   b_\mu \Bigg(  1-  \left(1+r[\Ph]\right)     \left\{  \sum_{u=1}^{M_S} r_{\mu u}^S  \tilde{A}^S_u A_u^{0S}+  \sum_{v=1}^{M_A} r_{\mu v}   \tilde{A}^T_v A_v^{0T}      \right\} \frac{\phi(M)}{M}+O\left(\frac{\phi^2(M)}{M^2} \right)    \Bigg  )  \nonumber
\\
&=& b_\mu +O\left(\phi(M)/M\right)
\end{eqnarray}
 Inserting this into equation  [\ref{eq:multi-Ab-Ag-MF-r[P]-2}] leads us for $\ab\neq\nullv$ to
\begin{eqnarray}
r \left[\Ph\right]&=&\frac{  r P_0 }{    \frac{r}{M}\sum_{\mu=1}^M \left\{ \sum_{u=1}^{M_S} r_{\mu u}^S  \tilde{A}^S_u A_u^{0S}    + \sum_{v=1}^{M_A} r_{\mu v}  \tilde{A}^T_v A_v^{0T}\right\}   \left\{ b_\mu +O\left(\frac{\phi(M)}{M} \right)  \right\}  +  \frac{1}{\phi(M)}   }\label{eq:multi-Ab-Ag-MF-r[P]-large-M}\\
&=&  \frac{ P_0 }{ \sum_{u=1}^{M_S}  B^S_u(\ab) \tilde{A}^S_u A_u^{0S}    +\sum_{v=1}^{M_A}   B_v(\ab)   \tilde{A}^T_v A_v^{0T}   +  \frac{1}{r\phi(M)}  +O\left(\frac{\phi(M)}{rM} \right)  }\nonumber\\
&=&  \frac{ P_0 }{ \sum_{u=1}^{M_S}  B^S_u(\ab) \tilde{A}^S_u A_u^{0S}    +\sum_{v=1}^{M_A}   B_v(\ab)   \tilde{A}^T_v A_v^{0T}    }+ O\left(\frac{1}{r\phi(M)} \right)  \nonumber
%
%
%
\end{eqnarray}
Here we have  defined the following two macroscopic observables:
\begin{eqnarray}
B^T_v(\ab)=  \frac{1}{M}\sum_{\mu=1}^M  r_{\mu v} \, b_\mu,~~~~~~~~
\label{def:B-2}
 B^S_u(\ab)=\frac{1}{M}\sum_{\mu=1}^M   r_{\mu u}^S   b_\mu\nonumber. 
\end{eqnarray}
Finally, for the normalised self-Ag $\tilde{A}^S_u=A^S_u/ A^S_u(\nullv) $ and the normalised target Ag $ \tilde{A}^T_v=A^T_v/ A^T_v(\nullv) $ we proceed in a similar way and obtain the  equations
\begin{eqnarray}
 \tilde{A}^S_u &=&   \frac{ 1 }{  1+\left(1+r[\Ph]\right)\sum_{\mu=1}^M r_{\mu u}^S  [{\I}_\mu  ] }   \label{eq:multi-Ab-Ag-MF-As-1}\\
&=&   \frac{  1  }{  1+\left(1+r[\Ph]\right)\frac{1}{M}\sum_{\mu=1}^M r_{\mu u}^S b_\mu +O\left(\frac{\phi(M)}{M} \right)   } \nonumber\\
&=&   \frac{  1  }{ \! 1\!+\left(\!1+\! \frac{ P_0 }{ \sum_{\tilde{u}=1}^{M_S}  B^S_{\tilde{u}}   (\ab) \tilde{A}^S_{\tilde{u}}  A_{\tilde{u}} ^{0S}    +\sum_{v=1}^{M_A}   B^T_v(\ab)   \tilde{A}^T_v A_v^{0T}    } \right)\! B^S_u(\ab)    +  O\!\left(\frac{1}{r\phi(M)} \right)   }    \label{eq:multi-Ab-Ag-MF-recursion-M-finite}\\
 \tilde{A}^T_v  &=&   \frac{  1  }{ \! 1\!+\left(\!1+\! \frac{ P_0 }{ \sum_{u=1}^{M_S}  B^S_u(\ab) \tilde{A}^S_u A_u^{0S}    +\sum_{{\tilde{v}} =1}^{M_A}   B^T_{\tilde{u}} (\ab)   \tilde{A}^T_{\tilde{u}}  A_{\tilde{u}}^{0T}    } \right)\! B^T_v(\ab)    +  O\!\left(\frac{1}{r\phi(M)} \right)   }    \nonumber
\end{eqnarray}
which for $M\rightarrow\infty$ is equivalent  to the system
\begin{eqnarray}
 A^S_u  =   \frac{  A^S_u(\nullv)  }{ \! 1\!+\left(\!1+\! \frac{ P(\nullv) }{ \sum_{{\tilde{u}} =1}^{M_S}  B^S_{\tilde{u}} (\ab) A^S_{\tilde{u}}      +\sum_{v=1}^{M_A}   B^T_v(\ab)   A^T_v    } \right)\! B^S_u(\ab)      },~~~~~~~~   \label{eq:multi-Ab-Ag-MF-recursion-M-infinite}
 A^T_v  =   \frac{  A^T_v(\nullv)  }{ \! 1\!+\left(\!1+\! \frac{ P(\nullv) }{ \sum_{u=1}^{M_S}  B^S_u(\ab) A^S_u     +\sum_{{\tilde{v}} =1}^{M_A}   B^T_{\tilde{v}} (\ab)   A^T_{\tilde{v}}      } \right)\! B^T_v(\ab)       }   
\end{eqnarray}
These expressions hold when $\ab\neq\nullv$. If  $\ab=\nullv$ we simply have   $A^S_u  =  A^S_u(\nullv) $ and $ A^T_v  =    A^T_v(\nullv)$. We note that the affinity parameter limit $r\rightarrow\infty$  and the repertoire size  limit $M\rightarrow\infty$ commute.  The meaning of the first limit  is that the forward rate of the reaction $AbAg + P\rightleftharpoons AbAgP$ in [\ref{eq:multi-Ab-Ag-reactions-2}] is much larger than the backward rate, i.e.  $K^+\gg K^-$. This limit enables us to use the present  equilibrium framework  to describe  also \emph{irreversible}  processes, such as Ag `removal'  reactions like  $AbAg + P\rightharpoonup AbAgP$~\cite{Gorban2011}.

\begin{figure}[t]
\centering
\includegraphics[width=0.4\textwidth]{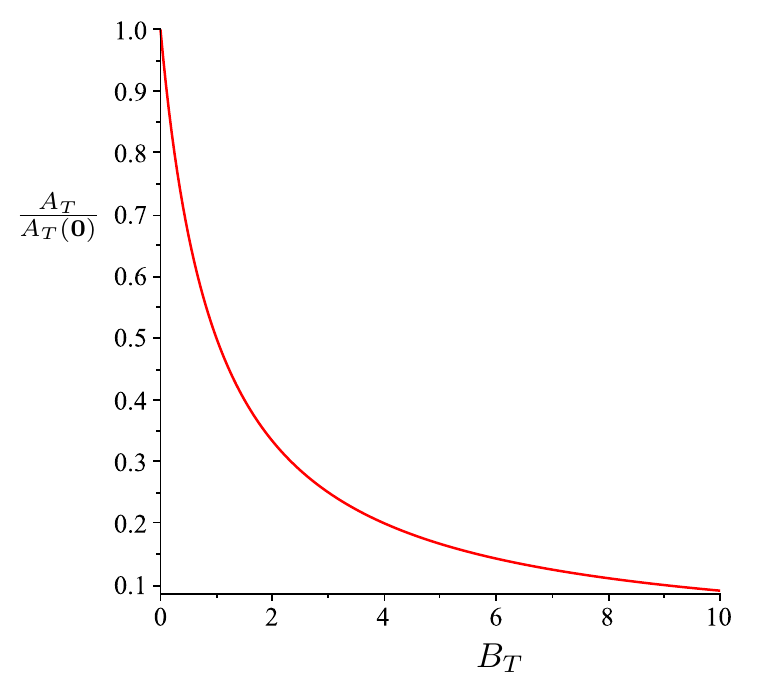}
\caption{Normalised free antigen concentration, $A_T/A_T(\nullv)$,  as a  function of the   average  of Ab concentrations $B_T(\ab)=M^{-1}\sum_{\mu=1}^M r_\mu b_\mu$. }
 \label{figure:AvsB} 
\end{figure}

The equations in [\ref{eq:multi-Ab-Ag-MF-recursion-M-infinite}] are functions of  the sum $y=\sum_{u=1}^{M_S}  B^S_u A^S_u     +\sum_{v =1}^{M_A}   B^T_{v}    A^T_{ v} $, which satisfies the recursive equation
\begin{eqnarray}
 y  &=&  \sum_{u=1}^{M_S}  \frac{  A^S_u(\nullv)B^S_u  }{ \! 1\!+\left(\!1+\! \frac{ P(\nullv) }{  y   } \right)\! B^S_u      } +  \sum_{v =1}^{M_A}   \frac{  A^T_v(\nullv) B^T_v   }{ \! 1\!+\left(\!1+\! \frac{ P(\nullv) }{  y      } \right)\! B^T_v       }  \label{eq:y-recursion}\\
  &=&    y\sum_{u=1}^{M_S}  \frac{  A^S_u(\nullv)B^S_u  }{ \left(1+   B^S_u\right)y +P(\nullv) B^S_u         } +  y\sum_{v =1}^{M_A}   \frac{  A^T_v(\nullv) B^T_v   }{  \left(1+  B^T_v\right)y  + P(\nullv) B^T_v         } \nonumber\\
    &=&    y\sum_{u=1}^{M_S}  \frac{  A^S_u(\nullv)B^S_u \prod_{\tilde{u}\neq u} \left[ \left(1+   B^S_{\tilde{u}}\right)y +P(\nullv) B^S_{\tilde{u}}  \right] }{ \prod_{\tilde{u}} \left[ \left(1+   B^S_{\tilde{u}}\right)y +P(\nullv) B^S_{\tilde{u}}  \right]       } + y\sum_{v =1}^{M_A}   \frac{  A^T_v(\nullv) B^T_v    \prod_{\tilde{v}\neq v} \left[ \left(1+   B^T_{\tilde{v}}\right)y +P(\nullv) B^T_{\tilde{v}}  \right]     }{    \prod_{\tilde{v}} \left[ \left(1+   B^T_{\tilde{v}}\right)y +P(\nullv) B^T_{\tilde{v}}  \right]           } \nonumber, 
\end{eqnarray}
where $B^S_u\equiv B^S_u(\ab)$ and $B^T_v\equiv B^T_v(\ab)$. The above identity follows directly from  the definition of  $y$ and   [\ref{eq:multi-Ab-Ag-MF-recursion-M-infinite}].  Thus $y$ is the   solution  of the following polynomial equation, of order $M_S+M_A$:
\begin{eqnarray}
&&\hspace*{-15mm}
\prod_{u=1}^{M_S} \left[ \left(1\!+\!   B^S_{  u  }\right)y \!+\!P(\nullv) B^S_{  u }  \right] \prod_{v =1}^{M_A}\left[ \left(1\!+\!   B^T_{  v }\right)y \!+\!P(\nullv) B^T_{v }  \right] 
\nonumber
\\
& =&  \sum_{u=1}^{M_S}   A^S_u(\nullv)B^S_u \prod_{\tilde{u}\neq u} \left[ \left(1\!+\!   B^S_{\tilde{u}}\right)y \!+\!P(\nullv) B^S_{\tilde{u}}  \right]  
+\sum_{v =1}^{M_A}    A^T_v(\nullv) B^T_v    \prod_{\tilde{v}\neq v} \left[ \left(1\!+\!   B^T_{\tilde{v}}\right)y \!+\!P(\nullv) B^T_{\tilde{v}}  \right] 
\label{eq:y-polynomial}
\end{eqnarray}
Let us assume that the relevant solution of [\ref{eq:y-polynomial}] is given by the function $\Phi \!\left(\B^T, \B^{S}  \right)$, where $\B^T=\left(B^T_1,\ldots,B^T_{M_A}\right)$ and  $\B^{S}=\left(B_1^S,\ldots,B^S_{M_S}\right)$, so that the solution of  the recursion [\ref{eq:multi-Ab-Ag-MF-recursion-M-infinite}] is given by 
\begin{eqnarray}
 A^S_u \left( \B^T, \B^{S}\right)  =   \frac{  A^S_u(\nullv)  \Phi \left(\B^T, \B^{S}  \right)  }{     \left(1+   B^S_u\right) \Phi \left(\B^T, \B^{S}  \right)        + P(\nullv)  B^S_u },~~~~~~~~   \label{eq:multi-Ab-Ag-MF-Ag-concentration}
 A^T_v \left( \B^T, \B^{S}\right)  =   \frac{  A^T_v(\nullv)  \Phi \left(\B^T, \B^{S}  \right) }{      \left(1+  B^T_v\right)   \Phi \left(\B^T, \B^{S}  \right) +   P(\nullv)     B^T_v                }   
\end{eqnarray}
and the   concentrations of (total) free self-Ag and target Ag are 
\begin{eqnarray}
 A_S\! \left( \ab\right)  =   \sum_{u=1}^{M_S} A^S_u \left( \B^T, \B^{S}\right)~~~~~~~~ 
  \label{eq:multi-Ab-Ag-MF-total-Ag}
 A_T\left( \ab\right)  =  \sum_{v=1}^{M_A} A^T_v \left( \B^T, \B^{S}\right).              
\end{eqnarray}
For $P(\nullv) =0$, i.e. in the absence of binding of Ag-Ab complexes to phagocytes, the above expressions simplify significantly to
\begin{eqnarray}
 A_S\! \left( \ab\right)  =   \sum_{u=1}^{M_S}  \frac{  A^S_u(\nullv)    }{   1+   B^S_u\left(\ab\right)}~~~~~~~~   \label{eq:multi-Ab-Ag-MF-total-Ag-no-P}
 A_T\left( \ab\right)  &=&  \sum_{v=1}^{M_A}   \frac{  A^T_v(\nullv)     }{ 1+  B^T_v\left(\ab\right)         },
\end{eqnarray}
so the concentration  of free Ag decreases with  increasing concentrations  of Abs.  In Figure    \ref{figure:AvsB}  we plot the (normalised) free target Ag concentration $A_T/A_T(\nullv) =1/(1+B(\ab))$ against  the average concentration of Abs $B_T(\ab)=M^{-1}\sum_{\mu=1}^Mr_\mu b_\mu$.
For $P(\nullv) >0$ we have to compute the function $ \Phi$  in [\ref{eq:multi-Ab-Ag-MF-Ag-concentration}].  Since, $\Phi$ is a solution of a polynomial of degree $M_A+M_S$  [\ref{eq:y-polynomial}],   this could be non-nontrivial.  But at least for  $M_A+M_S=2$ we can compute this function analytically.   Here  $ \Phi \left(\B^T, \B^{S}  \right)\equiv \Phi(B_T, B_S)$ is the  solution of the quadratic equation 
\begin{eqnarray}
0&=&\left( 1+ B_S \right) (1\! +\! B) y^2 \label{eq:quadratic}
+ \Big\{ B_T\left( 1\!+\! B_S \right) \left[P(\nullv) \!-\!A_T(\nullv)\right]  +   B_S  (1 \!+\! B_T) \left[P(\nullv)\! -\!A_S(\nullv)\right] \Big\} y  \nonumber\\
&&~~~~~~~ +~B_SB_T P(\nullv)   \left[P(\nullv)\!-\!A_S(\nullv)\!  -\!A_T(\nullv) \right].
\end{eqnarray}
Its determinant 
\begin{eqnarray}
D&=& \Big( B_T\left( 1\!+ \!B_S \right) \left[P(\nullv)\! -\!A_T(\nullv)\right]  +   B_S  (1 \!+\! B_T) \left[P(\nullv)\! -\!A_S(\nullv)\right] \Big)^2   \label{eq:D}\\
&&~~~~~~~~~~~~~~~~~~~~~~~~~~~~~ - 4 B_S B_T\left( 1\!+\! B_S \right) (1 \!+\! B_T) \, P(\nullv)   \left[P(\nullv)\!-\!A_S(\nullv)  -A_T(\nullv) \right]\nonumber
\end{eqnarray}
is positive when $A_T(\nullv)  + A_S(\nullv)\geq P(\nullv) $, in which case the  equation  has two real solutions. 
Only one of them is positive:
\begin{eqnarray}
\Phi(B, B_S)&=&\frac{ B_T\left[A_T(\nullv)\!-\!P(\nullv) \right]}{2\left( 1+ B_T \right)} +  \frac{ B_S\left[A_S(\nullv)\!-\!P(\nullv) \right] }{2\left( 1+ B_S \right)} \label{eq:Phi}\\
%
&&\hspace*{-0mm}
+~\Bigg\{\left(\frac{ B_T\left[A_T(\nullv)\! -\!P(\nullv) \right]}{2\left( 1+ B_T \right)} + \frac{ B_S\left[A_S(\nullv)\!-\!P(\nullv) \right] }{2\left( 1+ B_S \right)}\right)^2
+\frac{ B_T B_SP(\nullv) \left[A_T(\nullv) \!+\!A_S(\nullv)\!-\!P(\nullv) \right]}{\left( 1+ B_T \right)  \left( 1+ B_S \right)}   \Bigg\}^{\frac{1}{2}}.
\nonumber
\end{eqnarray}

\subsection{Bivalent Antibodies  reacting with  univalent target Antigen and self-Antigen} 
 In this section we show that in the regime of `weak' Abs, as considered in previous section, the amount of free Ag is not affected by the \emph{valency} of Abs~\cite{Janeway2012}. To this end it is sufficient only to consider the case of bivalent Abs interacting with univalent target Ag and self-Ag.  In particular we consider   $M$ different  bivalent   Abs, represented by the symbols  $\Y_\mu$ with $\mu\in\{1,\ldots,M\}$, forming complexes  with univalent target Ag, $\triangle$, and univalent  self-Ag, $\circ$.    The formation of complexes is modelled by the following chemical reactions:
  \begin{eqnarray}
  &&
\circ + \Y_\mu \overset{    {K_{\mu}^{S+}}   }{ \underset{    {K_{\mu}^{S-}}   }{\rightleftharpoons} }  \overset{\circ}{\Y_\mu}
~~~~~~~~~~~\label{eq:biv-Ab-univ-Ag-reactions}
\circ + \overset{\circ}{\Y_\mu} \overset{    {K_{\mu}^{S+}}   }{ \underset{    {K_{\mu}^{S-}}   }{\rightleftharpoons} }  \overset{\!\!\!\!\circ\,\circ}{\Y_\mu}~~~~~~~~~~~
\triangle + \Y_\mu \overset{    {K_{\mu}^{N+}}   }{ \underset{    {K_{\mu}^{N-}}   }{\rightleftharpoons} }  \overset{\triangle}{\Y_\mu}
\\
&&
\triangle + \overset{\triangle}{\Y_\mu} \overset{    {K_{\mu}^{N+}}   }{ \underset{    {K_{\mu}^{N-}}   }{\rightleftharpoons} }  \overset{\!\!\!\!\triangle\,\triangle}{\Y_\mu}
~~~~~~~~~
\circ +\overset{\triangle}{\Y_\mu}    \overset{    {K_{\mu}^{SN+}}   }{ \underset{    {K_{\mu}^{SN-}}   }{\rightleftharpoons} }          \overset{\!\!\!\!\circ\,\triangle}{\Y_\mu}
~~~~~~~~
\triangle +\overset{\circ}{\Y_\mu}           \overset{    {K_{\mu}^{SN+}}   }{ \underset{    {K_{\mu}^{SN-}}   }{\rightleftharpoons} }                  \overset{\!\!\!\!\triangle\,\circ}{\Y_\mu}                     
\end{eqnarray}
In chemical equilibrium, the concentrations of free self-Ag,  target Ag,  and Ab, which will be denoted, respectively, by the symbols $\left[\circ\right]$, $\left[\triangle\right]$ and $\left[\Y_\mu\right]$, are related to the concentrations of bound species $\overset{\circ}{\Y_\mu} $, $\overset{\!\!\!\!\circ\,\circ}{\Y_\mu}$, $\overset{\triangle}{\Y_\mu}$,  $\overset{\!\!\!\!\triangle\,\triangle}{\Y_\mu}$ and $ \overset{\!\!\!\!\triangle\,\circ}{\Y_\mu} $, which we denote, respectively, by the symbols $[\overset{\circ}{\Y_\mu} ]$, $[\overset{\!\!\!\!\circ\,\circ}{\Y_\mu}]$, $[\overset{\triangle}{\Y_\mu}]$,  $[\overset{\!\!\!\!\triangle\,\triangle}{\Y_\mu}]$ and $[ \overset{\!\!\!\!\triangle\,\circ}{\Y_\mu} ]$, via the affinities $r_{\mu}^S={K_{\mu}^{S+}}\! /{K_{\mu}^{S-}}$, $r_{\mu}^N={K^{N+}_{\mu }}\! /{K^{N-}_{\mu}}$ and $r_{\mu}^{SN}={K_{\mu}^{SN+}} \!/{K_{\mu}^{SN-}}$ via 
\begin{eqnarray}
r_{\mu}^S =\frac{  [\overset{\circ }{\Y_\mu}  ]     }{ \left[\circ\right]     \left[\Y_\mu\right] } =\frac{  [  \overset{\!\!\!\!\circ\,\circ}{\Y_\mu}    ]     }{ \left[\circ\right]    [  \overset{\circ }{\Y_\mu}   ] }\label{eq:biv-Ab-univ-Ag-equilibrium}
~~~~~~~~
r_{\mu}^N =  \frac{  [\overset{\triangle }{\Y_\mu}  ]     }{ \left[\triangle\right]     \left[\Y_\mu\right] } =\frac{  [  \overset{\!\!\!\!\triangle\,\triangle}{\Y_\mu}    ]     }{ \left[\triangle\right]    [  \overset{\triangle }{\Y_\mu}   ]  }~~~~~~~~
r_{\mu}^{SN}=   \frac{  [   \overset{\!\!\!\!\triangle\,\circ}{\Y_\mu}    ]     }{  [\triangle ]    [   \overset{\circ }{\Y_\mu}    ] }=   \frac{  [   \overset{\!\!\!\!\circ\,\triangle}{\Y_\mu}    ]     }{  [\circ ]    [   \overset{\triangle }{\Y_\mu}    ] }.
\end{eqnarray}
If the initial  concentrations of  species $\circ$, $\triangle$ and $\Y_\mu$   are, respectively,  given by  $\left[\circ\right]_0$, $\left[\triangle\right]_0$ and $\left[\Y_\mu\right]_0$   then,  because of  the mass conservation, we have 
\begin{eqnarray}
\left[\circ\right]_0&=&\left[\circ\right]+  \sum_{\mu=1}^M\,  [  \overset{\circ }{\Y_\mu}    ] + \sum_{\mu=1}^M\,[ \overset{\!\!\!\!\circ\,\triangle}{\Y_\mu}   ] + \sum_{\mu=1}^M\,[ \overset{\!\!\!\!\triangle\,\circ}{\Y_\mu}   ]  + 2\sum_{\mu=1}^M\,[ \overset{\!\!\!\!\circ\,\circ}{\Y_\mu}   ]  \label{eq:biv-Ab-univ-Ag-mass-1}\\
\left[\triangle\right]_0&=& \left[\triangle\right]   +  \sum_{\mu=1}^M\,  [  \overset{\triangle }{\Y_\mu}    ] + \sum_{\mu=1}^M\,[ \overset{\!\!\!\!\circ\,\triangle}{\Y_\mu}   ] + \sum_{\mu=1}^M\,[ \overset{\!\!\!\!\triangle\,\circ}{\Y_\mu}   ] + 2\sum_{\mu=1}^M\,[\, \overset{\!\!\!\!\triangle\,\triangle}{\Y_\mu}   ]  \nonumber\\
\left[\Y_\mu\right]_0&=&\left[\Y_\mu\right]+     [  \overset{\circ }{\Y_\mu}    ] +   [  \overset{\triangle }{\Y_\mu}    ] + [ \overset{\!\!\!\!\circ\,\triangle}{\Y_\mu}   ] + [ \overset{\!\!\!\!\triangle\,\circ}{\Y_\mu}   ] + [ \overset{\!\!\!\!\circ\,\circ}{\Y_\mu}   ]       + [\, \overset{\!\!\!\!\triangle\,\triangle}{\Y_\mu}   ].        \nonumber
\end{eqnarray}
Using the equilibrium relations [\ref{eq:biv-Ab-univ-Ag-equilibrium}] in the above three lines now gives us 
\begin{eqnarray}
\left[\circ\right] &=&\frac{ \left[\circ\right]_0   }{1+  \sum_{\mu=1}^M\left[ r_{\mu}^S      +  r_{\mu}^{SN}  \left[\triangle\right]  \left(r_{\mu}^{S} +r_{\mu}^{N}\right )+   2{r_{\mu}^S}^2  \left[\circ\right]      \right] \left[\Y_\mu\right]  }\label{eq:biv-Ab-univ-Ag-mass-2}\\
 \left[\triangle\right]  &=&\frac{  \left[\triangle\right]_0}{1  +  \sum_{\mu=1}^M\left[ r_{\mu}^N       +  r_{\mu}^{SN}  \left[\circ\right]    \left(r_{\mu}^{S} +r_{\mu}^{N}\right )+  2 {r_{\mu}^N}^2  \left[\triangle\right] \right]   \left[\Y_\mu\right] }\nonumber,
\end{eqnarray}
where
\begin{eqnarray}
\left[\Y_\mu\right]   &=&  \frac{\left[\Y_\mu\right]_0}  {1+     r_{\mu}^S  \left[\circ\right]      +     r_{\mu}^N  \left[\triangle\right]    + r_{\mu}^{SN}  \left[\circ\right]  \left[\triangle\right]        \left\{r_{\mu}^{S} +r_{\mu}^{N}\right \} +     {r_{\mu}^S}^2  \left[\circ\right]^2        +     {r_{\mu}^N}^2  \left[\triangle\right]^2      }.     \label{eq:Y-concentr}
\end{eqnarray}
Finally, with the notation $A_S= \left[\circ\right] $, $A_S(\nullv) =  \left[\circ\right]_0$, $ A_T= \left[\triangle\right]$, $ A_T(\nullv)=  \left[\triangle\right]_0$ and $b_\mu=\left[\Y_\mu\right]_0$,  we obtain  the recursive equations 
\begin{eqnarray}
A_S&=&\frac{ A_S(\nullv) }{1+  \sum_{\mu=1}^M  \frac{ \left[ r_{\mu}^S      +  r_{\mu}^{SN}   \left(r_{\mu}^{S} +r_{\mu}^{N}\right )A_T+   2{r_{\mu}^S}^2  A_S     \right]\, b_\mu } { 1+      r_{\mu}^S  A_S     +     r_{\mu}^N  A_T   +  r_{\mu}^{SN}  \left\{r_{\mu}^{S} +r_{\mu}^{N}\right \}A_S A_T +     {r_{\mu}^S}^2  A_S^2        +    {r_{\mu}^N}^2  A_T^2}          }\label{eq:biv-Ab-univ-Ag-recursion}\\
 A_T &=&\frac{  A_T(\nullv)}{1  +  \sum_{\mu=1}^M   \frac{ \left[ r_{\mu}^N       +  r_{\mu}^{SN}     \left(r_{\mu}^{S} +r_{\mu}^{N}\right )A_S+  2 {r_{\mu}^N}^2  A_T   \right]\,      b_\mu}{ 1+      r_{\mu}^S  A_S     +     r_{\mu}^N  A_T   +  r_{\mu}^{SN}  \left\{r_{\mu}^{S} +r_{\mu}^{N}\right \}A_S A_T +     {r_{\mu}^S}^2  A_S^2        +    {r_{\mu}^N}^2  A_T^2   }     }\nonumber.
\end{eqnarray}
%
 Now let us redefine $r_{\mu}=r_{\mu}/M$,  $r_{\mu}^S=r_{\mu}^S/M$ and $r_{\mu}^{SN}=r_{\mu}^{SN}/M$, and consider the relevant term in our expression for $A_S$:
\begin{eqnarray}
&&\hspace*{-10mm} \frac{ \left[ r_{\mu}^S      +  r_{\mu}^{SN}   \left(r_{\mu}^{S} +r_{\mu}^{N}\right )A_T+   2{r_{\mu}^S}^2  A_S \right] b_\mu  } { 1+      r_{\mu}^S  A_S     +     r_{\mu}^N  A_T   +  r_{\mu}^{SN}  \left\{r_{\mu}^{S} +r_{\mu}^{N}\right \}A_S A_T +     {r_{\mu}^S}^2  A_S^2        +    {r_{\mu}^N}^2  A_T^2}          \label{eq:biv-Ab-univ-Ag-MF-limit} \\
&=& \frac{ \frac{r_{\mu}^S}{M} b_\mu    +   \left[r_{\mu}^{SN}   \left(r_{\mu}^{S} +r_{\mu}^{N}\right )A_T+   2{r_{\mu}^S}^2  A_S \right]   \frac{b_\mu}{M^2} } { 1+     \left[ r_{\mu}^S  A_S    +     r_{\mu}^N  A_T \right] \frac{1}{M}  +  \left[r_{\mu}^{SN}  \left\{r_{\mu}^{S} +r_{\mu}^{N}\right \} A_S A_T +     {r_{\mu}^S}^2  A^2_S       +    {r_{\mu}^N}^2   A_T^2     \right]  \frac{1 }{M^2}     }      \nonumber\\
&=& \frac{r_{\mu}^S\, b_\mu}{M}  +O\left(\phi^2(M)/M^2\right)\nonumber
\end{eqnarray}
Here we assumed that $A_S, A\propto \phi(M)$, where  $\phi(M)= o(M)$.  The same argument applies to the corresponding term in the equation for $A_T$,  giving us $r_{\mu} b_\mu/M +O\left(\phi^2(M)/M^2\right)$ and hence 
\begin{eqnarray}
A_S(\ab)=\frac{ A_S(\nullv) }{1+  \frac{1}{M} \sum_{\mu=1}^M  r_{\mu}^S  b_\mu  }\label{eq:biv-Ab-univ-Ag-MF}
~~~~~~~~
 A_T (\ab)=\frac{  A_T(\nullv)}{1  +   \frac{1}{M}\sum_{\mu=1}^M   r_{\mu}^N    b_\mu }
\end{eqnarray}
for $M\rightarrow\infty$,  so we recover the result [\ref{eq:multi-Ab-Ag-MF-total-Ag-no-P}] for univalent Abs interacting with two types of Ag. The above argument easily generalises to include multiple univalent Ags and binding of Ag-Ab complexes. 

\section{Analysis of Antibody Dynamics\label{section:Ab-dynamics}}
In this section we study the Euler-Lagrange equation
\begin{eqnarray}
\Lambda_\mu\frac{\rmd^2}{\rmd t^2}{b}_\mu&=&-\frac{\partial}{\partial b_\mu}   \left[A_T(\ab) - \gamma A_S(\ab) \right]     \label{eq:E-L-quadr-2}, 
\end{eqnarray}
where $\Lambda_\mu\geq0$ and $\gamma\geq0$,  with the `energy' functions $A_T(\ab)$  and $ A_S(\ab)$  derived  in section \ref{ssection:univalent-Ab-Ag-reactions}.

\subsection{Binding  of univalent Antigens by univalent  Antibodies in the presence of univalent self-Antigens\label{ssection:multi-Ag-Ab-binding}}
Let us define the total  potential `energy' 
\begin{eqnarray}
 A_{\gamma} \left( \B^T, \B^{S}\right)&=&  A_T\left( \B^T, \B^{S}  \right) - \gamma A_S\left(  \B^T, \B^{S}    \right) 
  \label{def:A-gamma},
\end{eqnarray}
where $A_T\left( \B^T, \B^{S}  \right)\equiv A_T(\ab)$ and  $ A_S\left(  \B^T, \B^{S}    \right)\equiv  A_S(\ab)$, with   $A_T(\ab)$ and $A_S(\ab)$ as defined in [\ref{eq:multi-Ab-Ag-MF-total-Ag}],  and we consider equation  [\ref{eq:E-L-quadr-2}] for this energy function:
\begin{eqnarray}
\Lambda_\mu\frac{\rmd^2}{\rmd t^2}{b}_\mu&=&-\frac{\partial}{\partial b_\mu}   A_{\gamma} \left( \B^T, \B^{S}\right)   \label{eq:multi-Ab-Ag-MF-E-L-0}
~=~- \sum_{k=1}^{M_A}\frac{\partial  A_\gamma}{\partial B^T_{k}}  \frac{\partial B^T_{k}}{\partial b_\mu}  -  \sum_{\ell=1}^{M_S} \frac{\partial  A_\gamma}{\partial B^S_{\ell}   }   \frac{\partial  B^S_{\ell}    }{\partial b_\mu} \nonumber\\
&=& -  \sum_{k=1}^{M_A}\frac{\partial  A_\gamma}{\partial B^T_{k}}  \frac{  r_{\mu k} }{  M } -  \sum_{\ell=1}^{M_S} \frac{\partial  A_\gamma}{\partial B^S_{\ell}   }      \frac{r^S_{\mu\ell}    }{M} \nonumber
\end{eqnarray}
Assuming that   $\Lambda_\mu= \lambda_\mu \phi(M)  /M$, where $\phi(M)= o(M)$,  and using definition [\ref{def:B-2}]  above, allows us to derive the following equations for the set of  macroscopic observables $ \B^T$ and $\B^{S}$:
\begin{eqnarray}
\frac{\rmd^2}{\rmd t^2}{B}_{v}^T&=&- \sum_{k=1}^{M_A} 
(\rv_{v} \!\cdot\rv_{k}) \frac{\partial    }{\partial B_{k}^T}   A_{\gamma} \left( \B^T\!, \B^{S}\right)
 -  \sum_{\ell=1}^{M_S}  (\rv_{v} \!\cdot\rv_{\ell}^S)\frac{\partial  }{\partial B^S_{\ell}   }  A_{\gamma} \left( \B^T\!, \B^{S}\right)
   \label{eq:multi-dBdt}\\
\frac{\rmd^2}{\rmd t^2}{B}_{u}^S&=& - \sum_{k=1}^{M_A}
(\rv_{u}^S\!\cdot\rv_{k})  \frac{\partial }{\partial B_{k}^T} A_{\gamma} \left( \B^T\!, \B^{S}\right)
 - \sum_{\ell=1}^{M_S}  (\rv_{u}^S\!\cdot \rv^S_{\ell}) \frac{\partial }{\partial B^S_{\ell}   }  A_{\gamma} \left( \B^T\!, \B^{S}\right)
 \label{eq:multi-dBdt2}
\end{eqnarray}
with the short-hand $\xv \cdot\yv= M^{-1}\sum_{\mu=1}^M \lambda_\mu^{-1}    x_{\mu}   y_{\mu}$, with associated inner product norm $\vert \xv\vert=\sqrt{\xv \cdot\xv}$. 

In the special simplified case 
where each  Ab $\mu$ interacts with only \emph{one} type of Ag, we will have $\rv_{v} \cdot\rv_{k} =0$ if $v\neq k$,   $\rv_{v} \cdot \rv_{\ell}^S=0$, etc., and  the system of equations [\ref{eq:multi-dBdt}] simplifies to 
\begin{eqnarray}
\frac{1}{ \vert\rv_{v} \vert^2}\frac{\rmd^2}{\rmd t^2}{B}_{v}=- \frac{\partial    }{\partial B_{v}}   A_{\gamma} \!\left( \B, \B^{S}\right)
   \label{eq:multi-dBdt-reduced}
   ~~~~~~~~
\frac{1}{ \vert\rv_{u}^S\vert^2}\frac{\rmd^2}{\rmd t^2}{B}_{u}^S= 
 -   \frac{\partial }{\partial B^S_{u}   }  A_{\gamma}\! \left( \B, \B^{S}\right).
   \nonumber
\end{eqnarray}
We note that the above simplified  macroscopic dynamics is \emph{conservative}~\cite{Arnold1989}, with the energy function
\begin{eqnarray}
E\!\left( \B^T\!, \frac{\rmd}{\rmd t}{\B^T}; \B^{S}\!,\frac{\rmd}{\rmd t}{\B}^{S}\right)&=&
\sum_{v=1}^{M_A}\frac{1}{2  \vert\rv_{v} \vert^2}\Big(\frac{\rmd B^T_{v}}{\rmd t}\Big)^{\!2} 
+ \sum_{u=1}^{M_S}\frac{1}{2    \vert\rv_{u}^S\vert^2}  \Big(\frac{\rmd B^S_{u}}{\rmd t}\Big)^{\!2}   
+ A_{\gamma}\! \left( \B^T, \B^{S}\right)
   \label{eq:multi-Energy}
\end{eqnarray}
where the first two terms play the role of `kinetic' energies, and the third term is the `potential' energy.
The factors $1/\vert\rv_{v} \vert^2$ and $1/ \vert\rv_{u}^S\vert^2$ can be seen as  `masses'.  So     [\ref{eq:multi-dBdt-reduced}] describes  the motion~\cite{Arnold1989} of $M_A+M_S$ `particles', with  distinct masses,  in a potential field with potential energy    [\ref{def:A-gamma}]. 

\begin{figure}[t]
\centering
\includegraphics[width=0.4\textwidth]{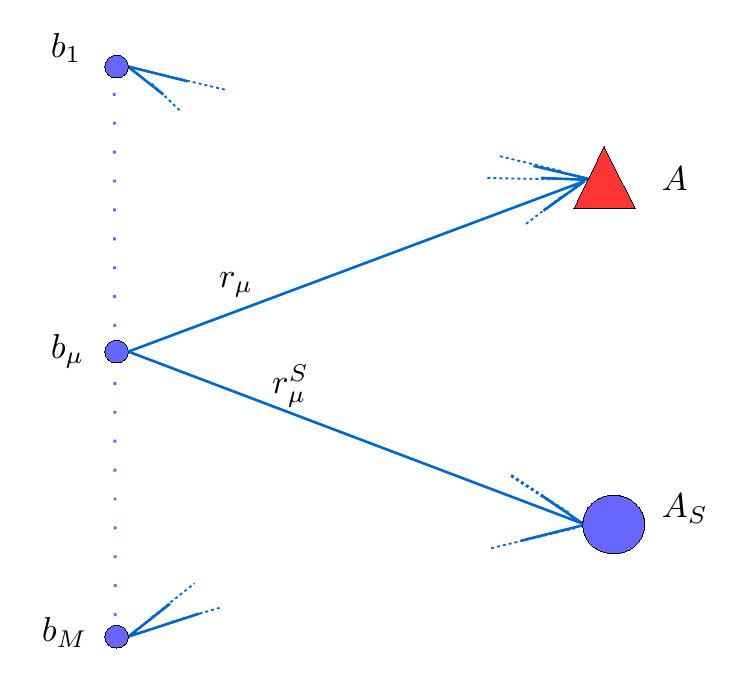}
\caption{ Network representation of  $M$ different populations of  univalent Abs (small blue circles) interacting with populations of  univalent target Ag (red triangle) and self-Ag (large blue circle).    Ab  $\mu$  is interacting with the  target Ag  and self-Ag with,  strengths (affinities) $r_\mu$ and   $r_\mu^S$, respectively.   The Ags are  interacting with   all Abs.   }
  \label{figure:Ab-Ag-Network} 
\end{figure}

Let us now assume that the numbers of target   and self Ags are equal, i.e.  $M_A=M_S$, and  that each Ab $\mu$ {simultaneously} interacts  with two types of Ag, one target and one self (see Figure  \ref{figure:Ab-Ag-Network} for $M_A=M_S=1$).    Then  the affinity vectors $\rv_{v}$ and $\rv_{u}^S$ satisfy  the \emph{orthogonality} conditions  $ \rv_{v} \cdot\rv_{k} =0$  if $k\neq v$ and  $\rv_{u}^S \cdot\rv_{\ell}^S=0$ if $\ell\neq u$, i.e. each row in the \emph{affinity matrices} $\Rmatrix^T=(\rv_{1},\ldots, \rv_{M_A})$ and $\Rmatrix^S=(\rv^S_{1},\ldots, \rv^S_{M_A})$ has exactly one positive component.
 Also  $\rv_{v} \cdot\rv_{\ell}^S =0$ if $\ell\neq u$, so, up to a permutation of columns, the matrices  $\Rmatrix^T$ and  $\Rmatrix^S$ are the same.   Our equations then simplify to
\begin{eqnarray}
\frac{\rmd^2}{\rmd t^2}{B}_{v}^T&=&- \frac{\partial    }{\partial B_{v}^T}   A_{\gamma} \!\left( \B^T\!, \B^{S}\right)
 \vert\rv_{v} \vert^2
 -   \frac{\partial  }{\partial B^S_{u}   }  A_{\gamma} \left( \B^T\!, \B^{S}\right)
(\rv_{v} \!\cdot\rv_{u}^S)
   \label{eq:multi-dBdt-reduced-2}\\
\frac{\rmd^2}{\rmd t^2}{B}_{u}^S&=& - \frac{\partial }{\partial B_{v}^T} A_{\gamma} \left( \B^T\!, \B^{S}\right)
 (\rv_{u}^S\!\cdot\rv_{v}) 
 -   \frac{\partial }{\partial B^S_{u}   }  A_{\gamma} \left( \B^T\!, \B^{S}\right)
  \vert\rv_{u}^S\vert^2.
\end{eqnarray}
Assuming that the above system is in `mechanical'  \emph{equilibrium}, $\rmd^2{B}^T_{v}/\rmd t^2=\rmd^2{B}_{u}^S/\rmd t^2=0$,  leads  us  to the two equalities 
\begin{eqnarray}
-\frac{\partial  A_{\gamma} \!\left( \B^T\!, \B^{S}\right)/\partial B^T_{v}}{\partial  A_{\gamma} \left( \B^T\!, \B^{S}\right)/\partial B^S_{u} }
 =
  \frac{(\rv_{v} \cdot\rv_{u}^S)}{\vert\rv_{v} \vert^2}  
   \label{eq:multi-equilibrium}
   ~~~~~~~~
-\frac{\partial A_{\gamma} \!\left( \B^T\!, \B^{S}\right)/\partial B^T_{v}}{\partial  A_{\gamma} \left( \B^T\!, \B^{S}\right)/\partial B^S_{u} }
=  
  \frac{\vert\rv_{u}^S\vert^2}{ (\rv_{u}^S\cdot\rv_{v})} 
\end{eqnarray}
and hence
\begin{eqnarray}
(\rv_{v} \cdot\rv_{u}^S)^2         &=&   \vert\rv_{u}^S\vert^2\vert\rv_{v} \vert^2 \label{eq:multi-equilibrium-affinities}.
\end{eqnarray}
We note that this will be true if and only if $\rv=\alpha(v,u)\rv_u^S$, for some $\alpha(v,u)>0$.
Using this in  [\ref{eq:multi-dBdt-reduced-2}] gives us the equations 
\begin{eqnarray}
\frac{\rmd^2}{\rmd t^2}{B}_{v}&=&- \frac{\partial    }{\partial B_{v}}   A_{\gamma} \!\left( \B, \B^{S}\right) \alpha^2 (v,u)  \vert\rv_{u}^S\vert^2
 -   \frac{\partial  }{\partial B^S_{u}   }  A_{\gamma} \left( \B, \B^{S}\right)
\alpha (v,u)  \vert\rv_{u}^S\vert^2  
   \label{eq:multi-dBdt-reduced-3}\\
\frac{\rmd^2}{\rmd t^2}{B}_{u}^S&=& - \frac{\partial }{\partial B_{v}} A_{\gamma} \left( \B, \B^{S}\right)
 \alpha (v,u)  \vert\rv_{u}^S\vert^2  
 -   \frac{\partial }{\partial B^S_{u}   }  A_{\gamma} \left( \B, \B^{S}\right)
  \vert\rv_{u}^S\vert^2 
   \nonumber
\end{eqnarray}
We note that $\alpha (v,u)$ generates a mapping $\rv_v=\alpha (v,u)\,\rv_{u}^S$ between the affinities  $\rv_v$ and $\rv_{u}^S$. Without loss of generality, we can always re-label the antibodies such that $u=v$, so that we only need $\alpha(v,v)\equiv \alpha(v)$. Equation  [\ref{eq:multi-dBdt-reduced-3}] can then be simplified to
\begin{eqnarray}
\frac{1}{\alpha(v)  \vert\rv_{v}^S\vert^2}\frac{\rmd^2}{\rmd t^2}{B}^T_{v}&=&- \frac{\partial    }{\partial B^T_{v}}   A_{\gamma} \!\left( \B^T\!, \B^{S}\right) \alpha(v) 
 -   \frac{\partial  }{\partial B^S_{v}   }  A_{\gamma} \left( \B^T\!, \B^{S}\right)
   \label{eq:multi-dBdt-reduced-4}\\
\frac{1}{ \vert\rv_{v}^S\vert^2}\frac{\rmd^2}{\rmd t^2}{B}_{v}^S&=& - \frac{\partial }{\partial B^T_{v}} A_{\gamma} \left( \B^T\!, \B^{S}\right)
 \alpha (v)   
 -   \frac{\partial }{\partial B^S_{v}   }  A_{\gamma} \left( \B^T\!, \B^{S}\right)
   \nonumber.
\end{eqnarray}
Furthermore, since now $B^T_{v}=\alpha(v) B^S_{v} $ the above reduces to the  single equation
\begin{eqnarray}
\frac{1}{ \vert\rv_{v}^S\vert^2}\frac{\rmd^2}{\rmd t^2}{B}_{v}^S&=& - \alpha (v)   \frac{\partial }{\partial B^T_{v}} A_{\gamma} \left( \B^T, \B^{S}\right)
 -   \frac{\partial }{\partial B^S_{v}   }  A_{\gamma} \left( \B^T, \B^{S}\right)
    \label{eq:multi-dBdt-reduced-5}, 
\end{eqnarray}
where the partial derivatives are evaluated  at $B^T_{v}=\alpha(v) B^S_{v}$. 

The macroscopic dynamics [\ref{eq:multi-dBdt-reduced-5}] is conservative when $P(\nullv)=0$. In   this case  the  potential energy [\ref{def:A-gamma}] is given by 
\begin{eqnarray}
 A_{\gamma} \left( \B^T\!, \B^{S}\right)&=& \sum_{v=1}^{M_A}\frac{  A^T_v(\nullv)  }{1+  B^T_v }   - \gamma    \sum_{u=1}^{M_S}\frac{ A^S_u(\nullv)  }{1+  B^S_u }   \label{eq:A-gamma-P-0}
\end{eqnarray}
and equation [\ref{eq:multi-dBdt-reduced-5}] reduces to 
\begin{eqnarray}
\frac{1}{ \vert\rv_{v}^S\vert^2}\frac{\rmd^2}{\rmd t^2}{B}_{v}^S&=& -\frac{\partial }{\partial B^S_{v}   } \left\{\frac{  A_v^{0T} }{1+  \alpha (v)  B^S_v  }
      - \gamma   \frac{ A^{0S}_v}{ 1+  B^S_v  }\right\}
    \label{eq:multi-dBdt-P-0-potential}, 
\end{eqnarray}
so this dynamics is  conservative,  with the energy 
\begin{eqnarray}
E_v\!\left(B^S_v, \frac{\rmd}{\rmd t}{B}_{v}^{S}\right)&=&\frac{1}{2 \vert\rv_{v}^S\vert^2}\Big(\frac{\rmd}{\rmd t}{B}_{v}^{S}\Big)^2 +\frac{  A_v^{0T} }{\left(1+  \alpha (v)  B^S_v  \right)}
      - \gamma   \frac{ A^{0S}_v}{\left( 1+  B^S_v \right) }
    \label{eq:multi-P-0-energy}, 
\end{eqnarray}
 describing  the  `motion'  a `particle'  of  `mass'  $1/\vert\rv_{v}^S\vert^2$ in a potential field.   If at time $t=0$ we are given the initial position  $B^S_v(0)$ and velocity  $(\rmd{B}^S_v/\rmd t)(0)$ of this particle,  then for all $t>0$ we have due to energy conservation:
\begin{eqnarray}
E_v\!\left(B^S_v, \frac{\rmd}{\rmd t}{B}_{v}^{S}\right) &=& E_v\!\left(B^S_v(0), (\frac{\rmd}{\rmd t}{B}_{v}^{S})(0)\right)\label{eq:multi-P-0-energy-t-0}.
\end{eqnarray}

\subsection{Binding  of univalent Antigen by univalent  Antibodies in the presence of univalent self-Antigen\label{ssection:Ag-Ab-binding}}
 The dynamics [\ref{eq:E-L-quadr-2}]  with the energy function [\ref{eq:A-gamma-P-0}]   can be solved in a  full detail when $M_A=M_S=1$ (see Figure  \ref{figure:Ab-Ag-Network}).  Here  the Euler-Lagrange equation is 
\begin{eqnarray}
\Lambda_\mu  \frac{\rmd^2}{\rmd t^2}{b}_\mu&=&-\frac{\partial}{\partial b_\mu}   \left[\frac{A_T(\nullv)}{1+ B_T(\ab)  } - \gamma \frac{A_S(\nullv)}{ 1+    B_S(\ab)  }     \right]\label{eq:E-L-MF-1}\\
&=&\frac{A_T(\nullv)}{\left(1+ B_T(\ab) \right)^2 }\frac{ r_\mu}{M} - \gamma \frac{A_S(\nullv)\, }{ \left(1+    B_S(\ab) \right)^2  } \frac{r^S_\mu}{M}    \nonumber,
\end{eqnarray}
where  $B_T(\ab)=M^{-1}\sum_{\nu=1}^Mr_\nu\, b_\nu(t)$ and $B_S(\ab)=M^{-1}\sum_{\nu=1}^M   r^S_\nu\, b_\nu$. The latter two macroscopic observables are governed  by the  equations 
\begin{eqnarray}
\frac{\rmd^2}{\rmd t^2}{B}=  \frac{A_T^0      \vert \rv\vert^2}{\left(1+  B_T \right)^2 } - \gamma \frac{A_S^0    (\rv\cdot\rv^S)}{ \left(1+     B_S \right)^2  }    \label{eq:MF-dynamics-2}
~~~~~~~~~~~
\frac{\rmd^2}{\rmd t^2}{B}_S=\frac{A_T^0    (\rv\cdot\rv^S)   }{\left(1+  B_T \right)^2 } - \gamma \frac{A_S^0   \vert \rv^S \vert^2}{ \left(1+    B_S \right)^2  },
\end{eqnarray} 
where $B_T\equiv B(\ab)$ and $ B_S\equiv  B_S(\ab)$, with initial conditions $\{(\rmd{B_T}/\rmd t)(0), (\rmd{B}_S/\rmd t)(0), B_T(0), B_S(0)\}$.  So the above equations are a special case of  [\ref{eq:multi-dBdt},\ref{eq:multi-dBdt2}].  Furthermore, the average concentration of Abs $\tilde{B}(\ab)=M^{-1}\sum_{\nu=1}^M b_\nu $  is governed by 
\begin{eqnarray}
\frac{\rmd^2}{\rmd t^2}{\tilde{B}}&=&  \frac{A_T^0    ( \rv\cdot\1)  }{\left(1+  B_T \right)^2 } - \gamma \frac{A_S^0     (\rv^S\!\cdot\1)}{ \left(1+     B_S \right)^2  }    \label{eq:MF-Ab-dynamics-2}.
\end{eqnarray}

The simplest case is that where each Ab is either self-reactive or non-self-reactive (never both), i.e. for  all $\mu$  either $r_\mu=0$ and  $r^S_\mu>0$ or $r_\mu>0$ and $r^S_\mu=0$. This implies  that $ ( \rv\cdot\rv^S)=0$ in [\ref{eq:MF-dynamics-2}],  giving us the two independent equations  
\begin{eqnarray}
\frac{\rmd^2}{\rmd t^2}{B}_T=  \frac{A_T^0     \vert \rv\vert^2    }{\left(1+  B_T \right)^2 }    \label{eq:MF-dynamics-Ab-dichotomy-2}
~~~~~~~~~~
\frac{\rmd^2}{\rmd t^2}{B}_S=- \gamma \frac{A_S^0 \vert \rv^S \vert^2  }{ \left(1+    B_S \right)^2  }.
\end{eqnarray}
We note that above is a special case of [\ref{eq:multi-dBdt-reduced}], so the dynamics of $B_T$ is conservative with the energy 
\begin{eqnarray}
E\!\left(B_T, \frac{\rmd}{\rmd t}{B_T}\right)&=&\frac{1}{2   \vert \rv\vert^2}\Big(\frac{\rmd}{\rmd t}{B_T}\Big)^2+ \frac{A_T^0       }{1+  B_T},
\end{eqnarray}
Since energy is conserved, one can then use the identity 
 $E\!\left(B_T, \rmd{B_T}/\rmd t\right)=E\left(B_T(0), (\rmd{B_T}/\rmd t)(0)\right)$  to obtain a simple equation for $\rmd{B}/\rmd t$.  For the initial conditions  $(\rmd{B_T}/\rmd t)(0)=B_T(0)=0$ this equation is given by 
\begin{eqnarray}
\frac{\rmd}{\rmd t}{B_T}&=&\sqrt{2 A_T^0    \vert \rv\vert^2   \frac{    B_T   }{1+  B_T} } \label{eq:MF-dynamics-dB/dt-2}. 
\end{eqnarray}
The function $\sqrt{B /(1\!+\!  B) }\in[0,1]$ is monotonic increasing and concave for $B\in [\,0,\infty)$.  Hence $B_T(t)$ is bounded from above by $\sqrt{2 A^0  \vert \rv\vert^2   }\, t $, saturating this upper bound as $t\rightarrow\infty$. Furthermore, the (normalised) amount    of antigen  $A_T/A(\nullv)=1/(1\!+\!  B_T(t)) $ is bounded from below by $1/(1\!+\!  \sqrt{2 A^0 \vert \rv\vert^2 )  \, t  }$. 
Also 
the dynamics of  $B_S$ in  [\ref{eq:MF-dynamics-Ab-dichotomy-2}] is conservative, with energy 
\begin{eqnarray}
E\!\left(B_S, \frac{\rmd}{\rmd t}{B}_S\right)&=&\frac{1}{2   \vert \rv^S\vert^2}\Big(\frac{\rmd}{\rmd t}{B}_S\Big)^2- \frac{\gamma A^0_S       }{1+  B_S},
\end{eqnarray}
and using $E\!\left(B_S, \rmd{B}_S/\rmd t\right)=E\!\left(B_S(0), (\rmd{B}_S/\rmd t)(0)\right)$, with initial conditions $(\rmd{B}_S/\rmd t)(0)=B_S(0)=0$,  gives us  the equation 
\begin{eqnarray}
\Big(\frac{\rmd}{\rmd t}{B}_S\Big)^2&=&-2\gamma A_S^0  \vert \rv^S\vert^2  \frac{  B_S    }{1+  B_S}  \label{eq:MF-dynamics-dBs/dt-2}
\end{eqnarray}
which for $\gamma>0$ has only  the trivial solution $B_S=0$. Values  $\gamma<0$ lead to self-antigen removal and hence are  not desirable.   

%
Further results for [\ref{eq:MF-dynamics-2}] can in equilibrium states, defined by $\rmd^2{B_T}/\rmd t^2=\rmd^2{B_S}/\rmd t^2=0$.  From these conditions we infer that $(\rv\cdot\rv^S)^2=\rv^2(\rv^S)^2$,   hence  $r_\mu=\alpha\, r^S_\mu$ for some  $\alpha>0$.  This, in return, via the definitions of $B_T$ and $B_S$, implies $B_T=\alpha B_S$ and hence  the system  [\ref{eq:MF-dynamics-2}] reduces to  a single  equation:
\begin{eqnarray}
\frac{\rmd^2}{\rmd t^2}{B}_S&=&A_S^0\vert \rv^S\vert^2   \left[\frac{\alpha\beta  }{\left(1+  \alpha B_S \right)^2 } - \frac{ \gamma}{ \left(1+    B_S \right)^2  }  \right]    \label{eq:Bs-ode},
\end{eqnarray}
where we defined $\beta=A_T^0/A_S^0$.  Furthermore, for equation  [\ref{eq:MF-Ab-dynamics-2}],  governing the average concentration of antibodies $\tilde{B}$, we obtain 
\begin{eqnarray}
\frac{\rmd^2}{\rmd t^2}{\tilde{B}}&=& A_S^0     (\rv^S\!\cdot\1 )    \!\left[\frac{\alpha\beta  }{\left(1+  \alpha B_S \right)^2 } - \frac{ \gamma}{ \left(1+    B_S \right)^2  }  \right]  
 \label{eq:MF-Ab-quadr-ode}.
\end{eqnarray}
Thus the two equations [\ref{eq:Bs-ode}] and [\ref{eq:MF-Ab-quadr-ode}] are related according to $\vert \rv^S\vert^2\rmd^2 \tilde{B}/\rmd t^2=(\rv^S\!\cdot\1)   \rmd^2{B}_S/\rmd t^2$, and hence 
\begin{eqnarray}
\tilde{B}&=&[(\rv^S\!\cdot\1)/ \vert \rv^S\vert^2 ] B_S\label{eq:MF-Ab-equality-2}.
\end{eqnarray}
The dynamics [\ref{eq:Bs-ode}] conserves the energy
\begin{eqnarray}
E\left(B_S,  \frac{\rmd}{\rmd t}{B}_S\right)&=&\frac{1}{2 \vert \rv^S\vert^2} \Big(\frac{\rmd}{\rmd t}{B}_S\Big)^2+A_S^0   \left[\frac{\beta  }{\left(1+  \alpha B_S \right) } -\frac{ \gamma}{ \left(1+    B_S\right)  }  \right]  
\end{eqnarray}
and we can use  $E\left(B_S,  \rmd{B}_S/\rmd t\right)=E\left(B_S(0), (\rmd{B}_S/\rmd t)(0)\right)$ to obtain 
\begin{eqnarray}
\frac{\rmd}{\rmd t}{B}_S&=&\sqrt{\Big(\frac{\rmd}{\rmd t}{B}_S(0)\Big)^2+2A_S^0 \vert \rv^S\vert^2   \Big[\frac{ \gamma}{1+    B_S}-\frac{\beta  }{1+  \alpha B_S}  - \Big( \frac{ \gamma}{1+    B_S(0)  }- \frac{\beta  }{1+  \alpha B_S(0) }  \Big) \Big]  } \label{eq:MF-velocity-B0}.
\end{eqnarray}
Let us assume that $B_S(0)=(\rmd{B}_S/\rmd t)(0)=0$ then this simplifies to
\begin{eqnarray}
\frac{\rmd}{\rmd t}{B}_S&=&\sqrt{2A_S^0\,  \vert \rv^S\vert^2   \Big(\frac{ \gamma}{1+    B_S}-\frac{\beta  }{1+  \alpha B_S} -\gamma+\beta \Big)} \label{eq:MF-velocity-2}. 
\end{eqnarray}
The argument of the square root above  is \emph{non-negative}  if  
\begin{eqnarray}
\alpha\beta/\gamma \geq  (1\!+\!  \alpha B_S)/(1\!+\!    B_S)  \label{eq:MF-ineq-1}, 
\end{eqnarray}
equivalently, if $\gamma/(1+    B_S)-\beta/(1+  \alpha B_S)  -\gamma+\beta \geq0$.  We note that for the $B_S=0$ and   $B_S=\infty$ this inequality reduces to $\alpha\beta\geq\gamma$ and  $\beta\geq\gamma$, respectively. The right hand side of [\ref{eq:MF-ineq-1}] is monotonically increasing  on the interval $B_S\in[\,0,\infty)$ when $\alpha>1$, and monotonically decreasing if  $\alpha<1$. Hence we need to satisfy 
$\beta \geq \gamma$ when $\alpha>1$, and $\alpha\beta\geq\gamma$ when $\alpha<1$. 
The RHS of [\ref{eq:MF-velocity-2}] is a monotonic increasing  function of $B_S$  when 
\begin{eqnarray}
\beta/\alpha>\gamma ~~{\rm for}~~  \alpha>1, ~~~~{\rm and}~~~~ 
\alpha\beta>\gamma ~~ {\rm for}~~ \alpha<1
\label{eq:MF-ineq-3}
\end{eqnarray}
Taking the limit  $B_S\rightarrow\infty$ in the right hand side of [\ref{eq:MF-velocity-2}] gives us 
\begin{eqnarray}
\frac{\rmd}{\rmd t}{B}_S&=&\sqrt{2A_S^0   \vert \rv^S\vert^2      \left (\beta-\gamma\right)   }  \label{eq:MF-large-B-1}
\end{eqnarray}
 and hence
\begin{eqnarray}
B_S(t)&=&\sqrt{2A_S^0  \vert \rv^S\vert^2  \left (\beta-\gamma\right) }\,  t+\mbox{const}.
\label{eq:MF-large-B-2}
\end{eqnarray}
If the above monotonicity condition [\ref{eq:MF-ineq-3}] is satisfied, then 
\begin{eqnarray}
B_S(t)&\leq& t/\tau\label{eq:MF-large-B-ub},
\end{eqnarray}
where $\tau$ is the time constant  
\begin{eqnarray}
\tau&=& 1/\sqrt{2A_S^0  \vert \rv^S\vert^2   \left (\beta-\gamma\right) }.
\end{eqnarray}
Furthermore,   for $\alpha>1$ the RHS of [\ref{eq:MF-velocity-2}]  has a has a maximum at 
\begin{eqnarray}
B_S^* &=&\frac{\alpha(\beta-\gamma)+(\alpha-1)\sqrt{\alpha\beta\gamma}}{\alpha(\alpha\gamma-\beta)}  \label{eq:MF-max-velocity},
\end{eqnarray}
when 
\begin{eqnarray}
\beta/\alpha<\gamma\leq\beta \label{eq:MF-ineq-4},
\end{eqnarray}
So here the time constant in  [\ref{eq:MF-large-B-ub}] is different, and given by 
\begin{eqnarray}
\tau&=&\frac{1}{ \sqrt{2A_S^0  \vert \rv^S\vert^2   \Big(\frac{ \gamma}{1+    B_S^*}-\frac{\beta  }{1+  \alpha B_S^*} +\beta-\gamma \Big)}} .
\end{eqnarray}

\begin{figure}[t]
\centering
\includegraphics[width=0.82\textwidth]{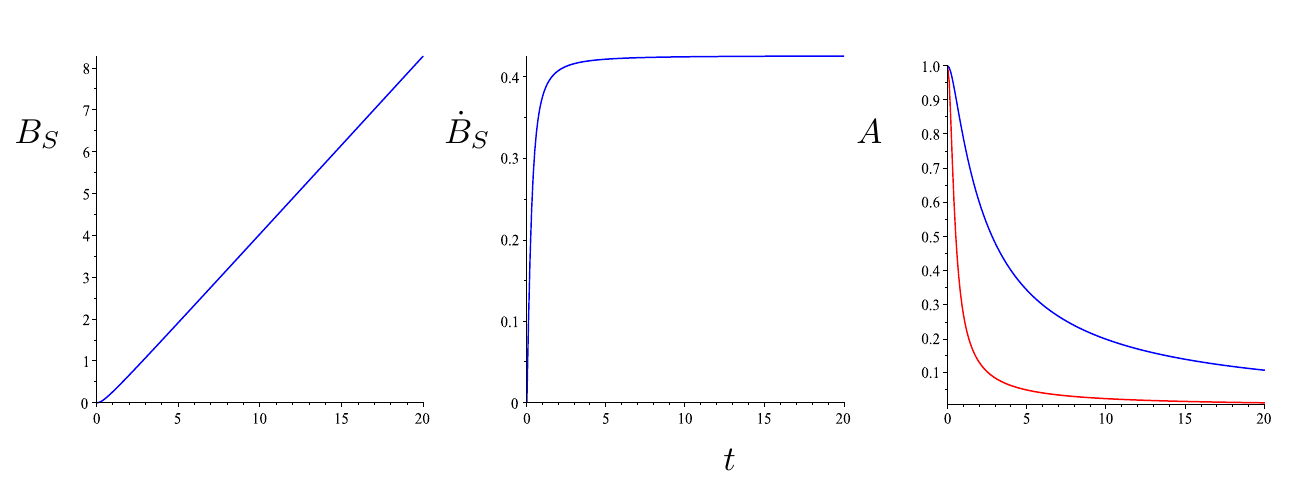}
\caption{The average Ab concentration, $B_S$,  and the rate $\dot{B}_S=\rmd B_S/\rmd t$ and  (normalised) Ag $A$ (top blue curve: self-Ag; bottom red curve: target Ag), shown  as functions of time $t$ for $A_S^0  \vert \rv^S\vert^2=10$,  $\alpha=10$, $\beta=0.01$ and $\gamma=0.0009$.}
\label{figure:Bvst-alpha-10-1} 
\end{figure}

\begin{figure}[h]
\centering
\includegraphics[width=0.82\textwidth]{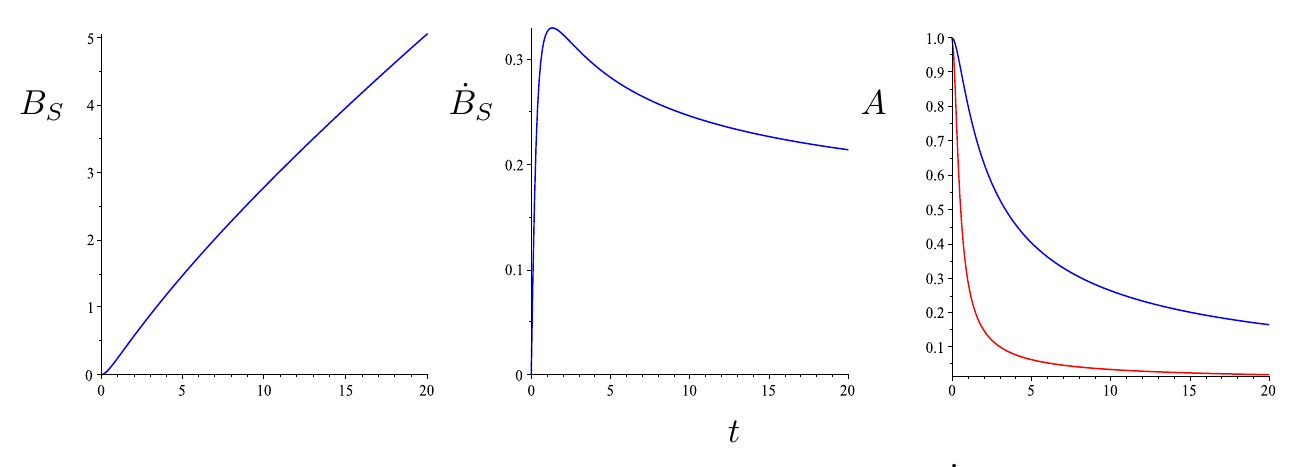}
\caption{The average Ab concentration, $B_S$,  and the rate $\dot{B}_S=\rmd B_S/\rmd t$ and  (normalised) Ag $A$  (top blue curve: self-Ag; bottom red curve: target Ag), shown  as functions of time $t$ for $A_S^0  \vert \rv^S\vert^2=10$,  $\alpha=10$, $\beta=0.01$ and $\gamma=0.009$.}
\label{figure:Bvst-alpha-10-2} 
\end{figure}

\begin{figure}[t]
\centering
\includegraphics[width=0.82\textwidth]{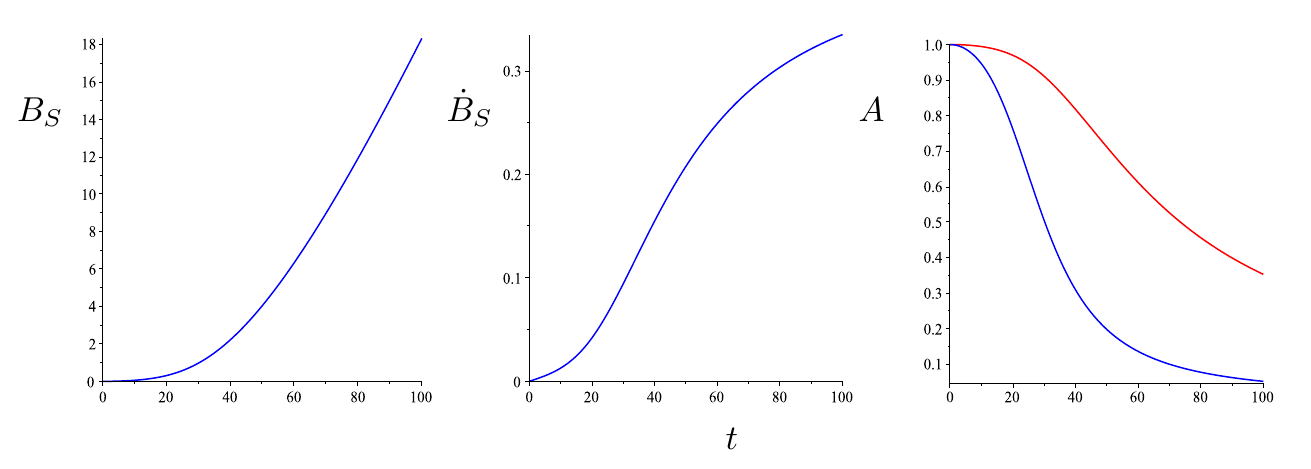}
\caption{ The average  Ab concentrations , $B_S$,  and the rate $\dot{B}_S$ and  (normalised) Ag $A$ (top blue curve: self-Ag; bottom red curve: target Ag), shown as functions of time $t$ for $A_S^0  \vert \rv^S\vert^2=10$,  $\alpha=1$, $\beta=0.01$ and $\gamma=0.009$. }
\label{figure:Bvst-alpha-1} 
\end{figure}

\begin{figure}[h]
\centering
\includegraphics[width=0.82\textwidth]{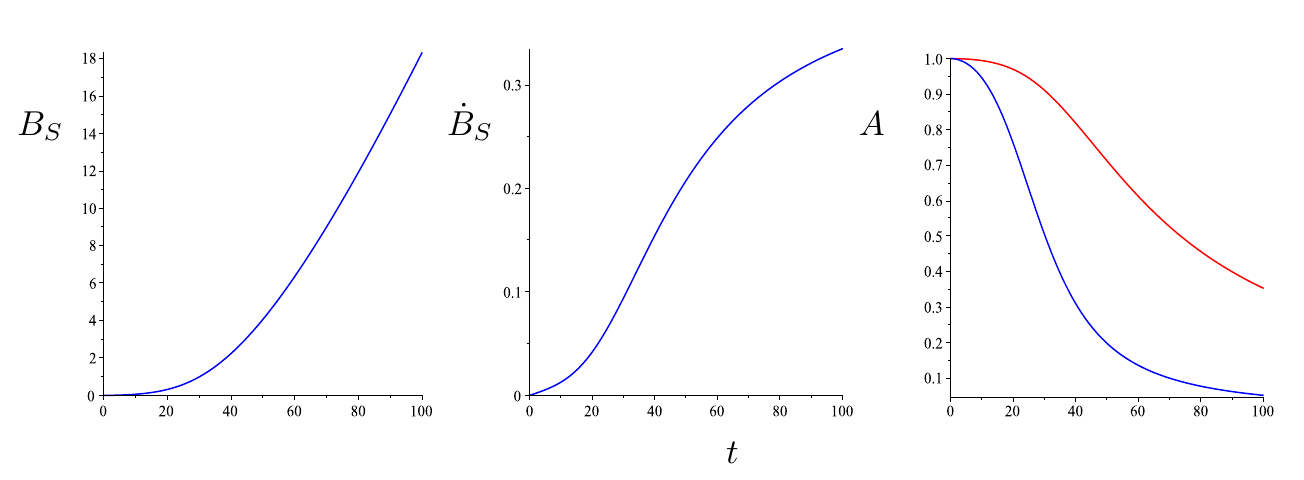}
\caption{The average Ab concentrations, $B_S$,  and the rate $\dot{B}_S$ and  (normalised) Ag $A$ (top blue curve: self-Ag;  bottom red curve: target Ag), shown   as functions of time $t$ for $A_S^0 \vert \rv^S\vert^2=10$,  $\alpha=0.1$, $\beta=0.01$ and $\gamma=0.009$.}
\label{figure:Bvst-alpha-01} 
\end{figure}

\begin{figure}[t]
\centering
\includegraphics[width=0.82\textwidth]{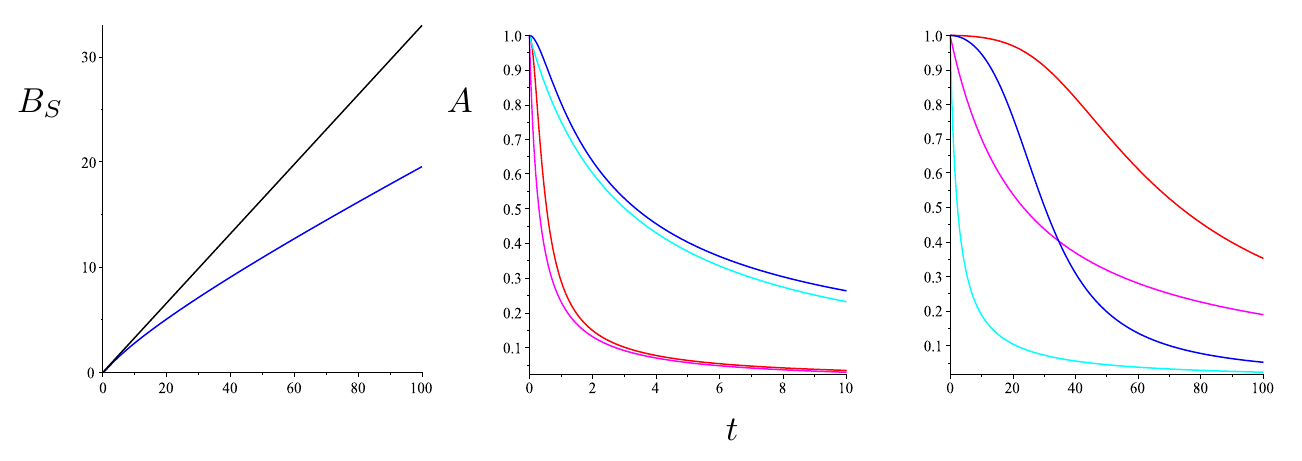}
\caption{ Left: The average  of Ab concentrations , $B_S$, (blue line)  and upper bound  (black line) as a function of time $t$ for $\alpha=10$ and $\gamma=0.009$. Middle: The (normalised) Ag,  $A$, (blue (cyan) curve  is  self-Ag and   red (magenta) curve is target Ag)  and the lower bound as a function of time $t$  for $\alpha=10$ and $\gamma=0.009$. Right: Ag and the lower bound (red (magenta) curve  is  target Ag and blue (cyan) curve is self-Ag) as a function of time $t$  for $\alpha=0.1$ and $\gamma=0.0009$.  The $\beta=0.01$  and  $A_S^0  \vert \rv^S\vert^2=10$.}
\label{figure:A-B-bounds} 
\end{figure}

We solve equation [\ref{eq:Bs-ode}] numerically in the regimes [\ref{eq:MF-ineq-3}] and [\ref{eq:MF-ineq-4}], for a given values of $\beta$ and $A_S^0  \vert \rv^S\vert^2$.  The solutions are plotted in Figures   \ref{figure:Bvst-alpha-10-1}--\ref{figure:Bvst-alpha-01}. Also we compare the upper bound [\ref{eq:MF-large-B-ub}]  with a typical solution of   [\ref{eq:Bs-ode}]  in Figure \ref{figure:A-B-bounds}.

 Let us now consider the normalised damage per unit of time 
\begin{eqnarray}
\delta_A(t_1-t_0)&=&\frac{1}{A(\ab(t_0))\left(t_1-t_0\right)}\int_{t_0}^{t_1}\!\rmd t^\prime~A(\ab({t^\prime})),
\label{def:damage-2}
\end{eqnarray}
where $0\leq\delta_A\leq1$,  and a similar integral  
\begin{eqnarray}
\delta_S(t_1-t_0)=\frac{1}{A_S(\ab(t_0))\left(t_1-t_0\right)}\int_{t_0}^{t_1}\!\rmd t^\prime~A_S(\ab({t^\prime})),
\label{def:self-damage-2}
\end{eqnarray}
where $0\leq\delta_S\leq1$, which defines  the (normalised)  self-damage  per unit of time $1-\delta_S$, where  $0\leq1-\delta_S\leq1$.  For  the scenario described by  the equation  [\ref{eq:Bs-ode}],  on the time interval $[0,t]$,  the above expressions  give us 
 \begin{eqnarray}
\delta_A(t)=\frac{1}{t}   \int_{0}^{t} \!\rmd t^\prime~       \frac{1}{1+  \alpha B_S({t^\prime})},            
\label{eq:damage}
~~~~~~~~~~
\delta_S(t)=\frac{1}{t} \int_{0}^{t} \!\rmd t^\prime~    \frac{1}{1+  B_S({t^\prime})}.
\end{eqnarray}
Since $1/(\!1+\!  \alpha B) $ decreases monotonically with $B$, from $B_S(t)< t/\tau$ we obtain  for the regime  [\ref{eq:MF-ineq-3}] the two lower bounds
 \begin{eqnarray}
\delta_A(t)\geq \delta_A^*(t)=\frac{\tau}{\alpha  t}\log\left(1+ \alpha \frac{t}{\tau}\right)\label{eq:damage-lb}
~~~~~~~~~~
\delta_S(t)  \geq\delta_S^*(t)  =\frac{\tau}{t}\log\left(1+ \frac{t}{\tau}\right),
\end{eqnarray}
with the time constant 
\begin{eqnarray}
\tau^{-1}&=&\sqrt{2A_S^0   \vert \rv^S\vert^2   \left (\beta-\gamma\right) } \label{def:MF-time-const-2},
\end{eqnarray}

 Let us consider the function $\delta^*(x)=x^{-1}\log(1+x)$ for $x\in(0,\infty)$.  Its derivative is ${\delta^*}^\prime(x)=[x- \left( 1+x \right)\log  \left( 1+x \right) ][\left( 1+x \right) x^2]$. Due to the inequality $\log  \left( 1+x \right)\geq1-(1+x)^{-1}$, this derivative is negative for any finite $x$, so $\delta^*(x)$ is a monotonic decreasing function with $\delta^*(x)\rightarrow1$ as $x\rightarrow0$ and  $\delta^*(x)\rightarrow0$ as $x\rightarrow\infty$.  Since the image of $\delta^*(x)$ is the interval $[0,1]$ the function $1-\delta^*(x)$ is monotonic increasing  on  the same domain.   It follows that $\delta_A^*(t)\rightarrow1$ as $t\rightarrow0$, implying that the (normalised) damage $\delta_A(t)\rightarrow1$ in this limit, and  $\delta_A^*(t)\rightarrow0$ as $t\rightarrow\infty$. Also  $1-\delta_S^*(t)  \rightarrow0$  as $t\rightarrow0$, implying that the self-damage  $1-\delta_S(t)  \rightarrow0$ in this limit, and $1-\delta_S^*(t)  \rightarrow1$ as $t\rightarrow\infty$. For $\alpha=1$ (where the strengths of antibody interaction with non-self and self are identical) we obtain $\delta_A^*=\delta_S^*$ and the damage $\delta_A^*$ (lower bound) is linearly related to the self-damage he self-damage $1-\delta_S^*$ (upper bound) via $\delta_A^*=1-(1-\delta_S^*)$. For $\alpha<1$ (where the strength of antibody interaction with self is greater than the interaction with  non-self) we obtain  $\delta_A^*>1-(1-\delta_S^*)$, so for a small reduction in  the damage $\delta_A^*$  we find a large increase in the  damage to self $1-\delta_S^*$.  For $\alpha>1$  (where the strength of antibody interaction with non-self is greater than the interaction with self)  we obtain  $\delta_A^*<1-(1-\delta_S^*)$, i.e. for a large reduction in $\delta_A^*$ we have a small  increase in $1-\delta_S^*$.

We (re-)label the antibodies such that $\lambda_1\leq \lambda_2\leq\cdots\leq\lambda_M$.  We define the mean and the variance of the binding strengths to self-antigen, $ m( \rv^S)= M^{-1}\sum_{\mu=1}^M r^S_\mu$ and $\sigma^2(\rv^S)=M^{-1}\sum_{\mu=1}^M(r^S_\mu)^2 -  (M^{-1}\sum_{\mu=1}^M r^S_\mu)^2$, and 
 consider  $ \vert \rv^S\vert^2= M^{-1}\sum_{\mu=1}^M \lambda_\mu^{-1}(r^S_\mu)^2$. We note that  for $\lambda_\mu=\lambda$:     
\begin{eqnarray}
 \lambda\,  \vert \rv^S\vert^2  &=& \sigma^2(\rv^S)+  m^2( \rv^S)    \label{eq:MF-2-nd-moment-2},
\end{eqnarray}
 Thus  the time constant $\tau$ is given by 

\begin{eqnarray}
1/ \tau(\lambda)&=& \sqrt{2A_S(\nullv)\,  \lambda^{-1}\left[\sigma^2(\rv^S)+  m^2( \rv^S)\right]     \left (\beta-\gamma\right) }  \label{eq:MF-time-const-1}
\end{eqnarray}
Second, the weighted average $M^{-1}\sum_{\mu=1}^M \lambda_\mu^{-1}\left(r^S_\mu\right)^2$, with $\lambda_\mu^{-1}\geq0$ for all $\mu$, is bounded from below by  $\lambda_{M}^{-1}M^{-1}\sum_{\mu=1}^M (r^S_\mu)^2$ and from above by  $\lambda_{1}^{-1}M^{-1}\sum_{\mu=1}^M(r^S_\mu)^2$. Hence the time constant in [\ref{def:MF-time-const-2}] is bounded according to
\begin{eqnarray}
\tau(\lambda_1)\leq \tau(\lambda)\leq \tau(\lambda_M)
\end{eqnarray}
This fact, in combination with 
the monotonicity of the $x^{-1}\log(1+x)$ as it appears in [\ref{eq:damage-lb}], gives us new lower bounds on the damage to non-self and the damage on self:
 \begin{eqnarray}
\delta_A(t)\geq \frac{ \tau\left(\lambda_{1}\right)}{\alpha\, t}\log\Big(1+ \alpha\, \frac{t}{\tau\left(\lambda_{1}\right)}\Big)\label{eq:damage-lbs}
~~~~~~~~
\delta_S(t)  \geq\frac{\tau\left(\lambda_{1}\right)}{t}\log\Big(1+  \frac{t}{\tau\left(\lambda_{1}\right)}   \Big)   
\end{eqnarray}
 We note that, since the time constant $\tau$ controls the speed of antigen removal,  see equation [\ref{eq:MF-velocity-2}], this speed   is a monotonic increasing function of   the  variance $\sigma^2(\rv^S)$ and the mean $m( \rv^S)$ of the vector of affinities $\rv^S$, i.e. of the antibody repertoire. Thus, having a repertoire with  a higher  variance facilitates  a more rapid  Ag removal.


\begin{thebibliography}{10}

\bibitem{Dunn-Walters2018}
Dunn-Walters D, Townsend C, Sinclair E, Stewart A (2018) Immunoglobulin gene
  analysis as a tool for investigating human immune responses.
\newblock {\em Immunological reviews} 284(1):132--147.

\bibitem{Dunn-Walters2016}
Dunn-Walters DK (2016) The ageing human b cell repertoire: a failure of
  selection?
\newblock {\em Clinical \& Experimental Immunology} 183(1):50--56.

\bibitem{Dondelinger2018}
Dondelinger M, et~al. (2018) Understanding the significance and implications of
  antibody numbering and antigen-binding surface/residue definition.
\newblock {\em Frontiers in immunology} 9.

\bibitem{Martin2016}
Martin VG, et~al. (2016) Transitional b cells in early human b cell
  development--time to revisit the paradigm?
\newblock {\em Frontiers in immunology} 7:546.

\bibitem{Childs2015}
Childs LM, Baskerville EB, Cobey S (2015) Trade-offs in antibody repertoires to
  complex antigens.
\newblock {\em Philosophical Transactions of the Royal Society B: Biological
  Sciences} 370(1676):20140245.

\bibitem{Wu2012}
Wu YCB, Kipling D, Dunn-Walters DK (2012) Age-related changes in human
  peripheral blood igh repertoire following vaccination.
\newblock {\em Frontiers in immunology} 3:193.

\bibitem{Poulsen2007}
Poulsen TR, Meijer PJ, Jensen A, Nielsen LS, Andersen PS (2007) Kinetic,
  affinity, and diversity limits of human polyclonal antibody responses against
  tetanus toxoid.
\newblock {\em The Journal of Immunology} 179(6):3841--3850.

\bibitem{Bannard2017}
Bannard O, Cyster JG (2017) Germinal centers: programmed for affinity
  maturation and antibody diversification.
\newblock {\em Current opinion in immunology} 45:21--30.

\bibitem{Ademokun2011}
Ademokun A, et~al. (2011) Vaccination-induced changes in human b-cell
  repertoire and pneumococcal igm and iga antibody at different ages.
\newblock {\em Aging cell} 10(6):922--930.

\bibitem{Goronzy2019}
Goronzy JJ, Weyand CM (2019) Mechanisms underlying t cell ageing.
\newblock {\em Nature Reviews Immunology} p.~1.

\bibitem{Martin2015}
Martin V, Wu YC, Kipling D, Dunn-Walters D (2015) Ageing of the b-cell
  repertoire.
\newblock {\em Philosophical Transactions of the Royal Society B: Biological
  Sciences} 370(1676):20140237.

\bibitem{Laffy2017}
Laffy JM, et~al. (2017) Promiscuous antibodies characterised by their
  physico-chemical properties: From sequence to structure and back.
\newblock {\em Progress in biophysics and molecular biology} 128:47--56.

\bibitem{Gibson2009}
Gibson KL, et~al. (2009) B-cell diversity decreases in old age and is
  correlated with poor health status.
\newblock {\em Aging cell} 8(1):18--25.

\bibitem{Kepler1993}
Kepler TB, Perelson AS (1993) Somatic hypermutation in b cells: An optimal
  control treatment.
\newblock {\em Journal of Theoretical Biology} 164(1):37 -- 64.

\bibitem{Agliari2011}
Agliari E, Barra A, Guerra F, Moauro F (2011) A thermodynamic perspective of
  immune capabilities.
\newblock {\em J. Theor. Biol.} 287:48--63.

\bibitem{Bartolucci2015}
Bartolucci S, Annibale A (2015) A dynamical model of the adaptive immune
  system: effects of cells promiscuity, antigens and b b interactions.
\newblock {\em J. Stat. Mech. Theory Exp.} 2015(8):P08017.

\bibitem{Mozeika2016}
Mozeika A, Coolen ACC (2016) Statistical mechanics of clonal expansion in
  lymphocyte networks modelled with slow and fast variables.
\newblock {\em J. Phys. A: Math. Theor.} 50(3):035602.

\bibitem{Theofilopoulos2017}
Theofilopoulos AN, Kono DH, Baccala R (2017) The multiple pathways to
  autoimmunity.
\newblock {\em Nature immunology} 18(7):716.

\bibitem{Perelson1979}
Perelson AS, Oster GF (1979) Theoretical studies of clonal selection: minimal
  antibody repertoire size and reliability of self-non-self discrimination.
\newblock {\em Journal of theoretical biology} 81(4):645--670.

\bibitem{DeBoer1993}
De~Boer RJ, Perelson AS (1993) How diverse should the immune system be?
\newblock {\em Proceedings of the Royal Society of London. Series B: Biological
  Sciences} 252(1335):171--175.

\bibitem{Mayer2015}
Mayer A, Balasubramanian V, Mora T, Walczak AM (2015) How a well-adapted immune
  system is organized.
\newblock {\em P. Natl. Acad. Sci. USA} 112(19):5950--5955.

\bibitem{Janeway2012}
Janeway C, Murphy KP, Travers P, Walport M (2012) {\em Janeway's
  Immunobiology}.
\newblock (Garland Science).

\bibitem{Arnold1989}
Arnold VI (1989) {\em Mathematical methods of classical mechanics}.
\newblock (Springer).

\bibitem{Gelfand2000}
Gelfand IM, Silverman RA, , et~al. (2000) {\em Calculus of variations}.
\newblock (Courier Corporation).

\bibitem{Yablonskii1991}
Yablonskii Gv, Bykov V, Elokhin V, Gorban A (1991) {\em Kinetic models of
  catalytic reactions}.
\newblock (Elsevier) Vol.{}~32.

\bibitem{Gorban2011}
Gorban A, Yablonsky G (2011) Extended detailed balance for systems with
  irreversible reactions.
\newblock {\em Chemical Engineering Science} 66(21):5388--5399.

\end{thebibliography}

\end{document}